\documentclass[10pt,journal,compsoc]{IEEEtran}
\ifCLASSOPTIONcompsoc
\else
\fi
\ifCLASSINFOpdf
\else
\fi
\usepackage{blindtext}

\usepackage{algorithm} 

\usepackage{algorithmic} 

\usepackage{graphicx}
\usepackage{amsfonts}
\usepackage{amsmath}
\usepackage{mathrsfs}
\usepackage{amssymb}
  \usepackage{graphicx}
\usepackage{setspace}
\usepackage{subfigure}
\usepackage{multirow}
\usepackage{url}

\begin{document}
%
\title{Distortion-Aware Concurrent Multipath Transfer for Mobile Video Streaming in Heterogeneous Wireless Networks}
%
%
%
%

\author{Jiyan~Wu,~\IEEEmembership{Student Member,~IEEE,} Bo~Cheng, Chau~Yuen,~\IEEEmembership{Senior Member,~IEEE,} Yanlei~Shang, and Junliang~Chen
\IEEEcompsocitemizethanks{\IEEEcompsocthanksitem Jiyan Wu, Bo Cheng, Yanlei Shang, and Junliang Chen are with the State Key Laboratory of Networking and Switching Technology, Beijing University of Posts and Telecommunications, Beijing 100876, P. R. China.\protect
\IEEEcompsocthanksitem Jiyan Wu and Chau Yuen are with the SUTD-MIT International Design Center, Singapore University of Technology and Design, 20 Dover Drive, Singapore 138682.\protect\\
E-mail: wujiyan@126.com, \{chengbo, shangyl, chjl\}@bupt.edu.cn, yuenchau@sutd.edu.sg.
}}

%
%

\markboth{IEEE TRANSACTIONS ON MOBILE COMPUTING,~VOL.~XX, NO.~X, 2014}%
{Shell \MakeLowercase{\textit{et al.}}: Bare Demo of IEEEtran.cls for Computer Society Journals}
%



\IEEEcompsoctitleabstractindextext{%
\begin{abstract}
The massive proliferation of wireless infrastructures with complementary characteristics prompts the bandwidth aggregation for Concurrent Multipath Transfer (CMT) over heterogeneous access networks. Stream Control Transmission Protocol (SCTP) is the standard transport-layer solution to enable CMT in multihomed communication environments. However, delivering high-quality streaming video with the existing CMT solutions still remains problematic due to the stringent QoS (Quality of Service) requirements and path asymmetry in heterogeneous wireless networks. In this paper, we advance the state of the art by introducing \emph{video distortion} into the decision process of multipath data transfer. The proposed Distortion-Aware Concurrent Multipath Transfer (CMT-DA) solution includes three phases: 1) per-path status estimation and congestion control; 2) quality-optimal video flow rate allocation; 3) delay and loss controlled data retransmission. The term `flow rate allocation' indicates dynamically picking appropriate access networks and assigning the transmission rates. We analytically formulate the data distribution over multiple communication paths to minimize the end-to-end video distortion and derive the solution based on the utility maximization theory. The performance of the proposed CMT-DA is evaluated through extensive semi-physical emulations in Exata involving H.264 video streaming. Experimental results show that CMT-DA outperforms the reference schemes in terms of video PSNR (Peak Signal-to-Noise Ratio), goodput, and inter-packet delay.
\end{abstract}


}

\maketitle

\IEEEdisplaynotcompsoctitleabstractindextext

%
\IEEEpeerreviewmaketitle

\section{Introduction}
\IEEEPARstart{D}uring the past few years, mobile video streaming service (e.g., Youtube [1], Hulu [2], online gaming, etc.) has become one of the ``killer applications'' and the video traffic headed for hand-held devices has experienced explosive growth. The latest market research conducted by Cisco company indicates that video streaming accounts for $53\%$ of the mobile Internet traffic in 2013 and will reach $69\%$ by the year 2018 [3]. In parallel, global mobile data is expected to increase $11$-fold in the next five years. Another ongoing trend feeding this tremendous growth is the popularity of powerful mobile terminals (e.g., smart phones and iPad), which facilitates individual users to access the Internet and watch videos from everywhere [4].
\begin{figure}[htbp]
\centering
 \includegraphics[width=0.5\textwidth,keepaspectratio]{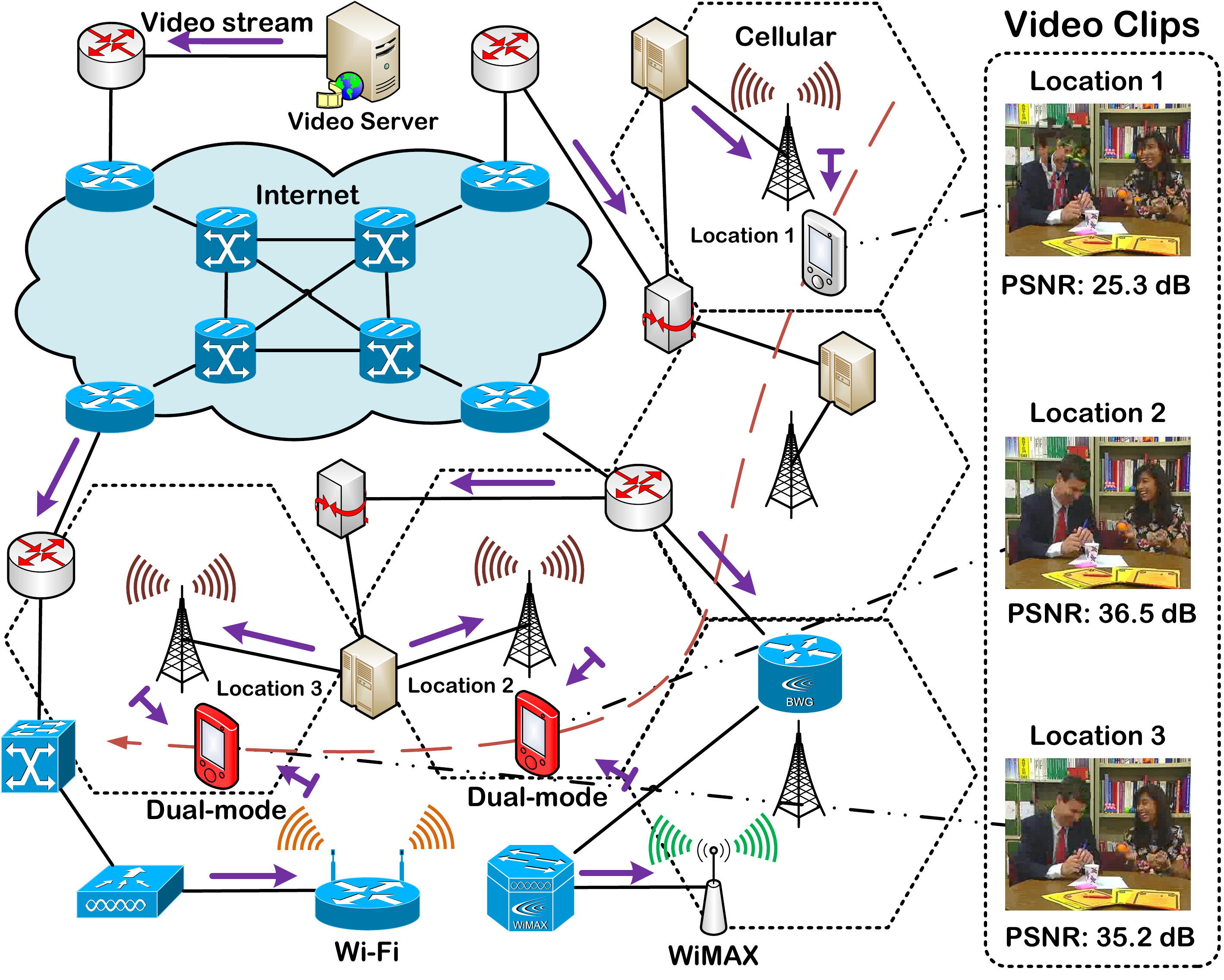}
 \caption{Illustration of exploiting concurrent multipath transfer for enhanced streaming video quality in heterogeneous wireless networks with multihomed clients.}\label{fig1}
\end{figure}

Despite the rapid advancements in network infrastructures, it is still challenging to deliver high-quality streaming video over wireless platforms [5-9][51]. On one hand, the Wi-Fi networks are limited in radio coverage and mobility support for individual users [5][6]; On the other hand, the cellular networks can well sustain the user mobility but their bandwidth is often inadequate to support the throughput-demanding video applications [7][8]. Although the 4G LTE and WiMAX can provide higher peak data rate and extended coverage, the available capacity will still be insufficient compared to the ever-growing video data traffic [9]. The complementary characteristics of heterogeneous access networks prompt the bandwidth aggregation for Concurrent Multipath Transfer (CMT) to enhance transmission throughput and reliability (see Fig. 1). With the emergency of multihomed/multinetwork terminals [10] (e.g., the Mushroom products [11]), CMT is considered to be a promising solution for supporting video streaming in future wireless networking.

The key research issue in multihomed video delivery over heterogeneous wireless networks must be effective integration of the limited channel resources available for providing adequate QoS (Quality of Service) [5-8][10]. Stream Control Transmission Protocol (SCTP) [12] is the standard transport-layer solution that exploits the multihoming feature to concurrently distribute data across multiple independent end-to-end paths [13]. Therefore, many CMT solutions [5][13][14] have been proposed to optimize the delay, throughput, or reliability performance for efficient data delivery. However, due to the special characteristics of streaming video, these network-level criteria cannot always improve the perceived media quality. For instance, a real-time video application encoded in Constant Bit Rate (CBR) may not effectively leverage the throughput gains since its streaming rate is typically fixed or bounded by the encoding schemes. In addition, involving a communication path with available bandwidth but long delay in the multipath video delivery may degrade the streaming video quality as the end-to-end distortion increases [50]. Consequently, leveraging the CMT for high-quality streaming video over heterogeneous wireless networks is largely unexplored.

In this paper, we investigate the problem by introducing \emph{video distortion} into the decision process of multipath data transfer over heterogeneous wireless networks. The proposed Distortion-Aware Concurrent Multipath Transfer (CMT-DA) solution is a transport-layer protocol and  includes three phases: 1) per-path status estimation and congestion control to exploit the available channel resources; 2) data flow rate allocation to minimize the end-to-end video distortion; 3) delay and loss constrained data retransmission for bandwidth conservation. The detailed descriptions of the proposed solution will be presented in Section 4. Specifically, the contributions of this paper can be summarized in the following.
\begin{itemize}
  \item An effective CMT solution that uses path status estimation, flow rate allocation, and retransmission control to optimize the real-time video quality in integrated heterogeneous wireless networks.
  \item A mathematical formulation of video data distribution over parallel communication paths to minimize the end-to-end distortion. The utility maximization theory is employed to derive the solution for optimal transmission rate assignment. 
  \item Extensive semi-physical emulations in Exata involving real-time H.264 video streaming. Experimental results show that: (1) CMT-DA increases the average video PSNR by up to $3.6$, $8.6$, and $11.3$ dB compared to the  CMT-QA [5], CMT-PF [14], and CMT [13], respectively. (2) CMT-DA improves the average goodput by up to $95$, $153$, and $195$ Kbps compared to the CMT-QA, CMT-PF, and CMT, respectively. (3) CMT-DA reduces the average inter-packet delay by up to $15.4$, $34.1$, and $51.1$ ms compared to the CMT-QA, CMT-PF, and CMT, respectively. (4) CMT-DA mitigates the effective loss rates by up to $3.7\%$, $7.2\%$, and $9.7\%$ compared to the CMT-QA, CMT-PF, and CMT, respectively.
\end{itemize}

The remainder of this paper is structured as follows. In Section 2, we review and discuss the related work. Section 3 presents the model and problem statement. Section 4 describes the design and solution of the proposed CMT-DA in detail. Performance evaluation is provided in Section 5 and conclusion remarks are given in Section 6. The basic notations used throughput this paper are listed in Table 1.

Our previous studies [33][38] on multihomed video delivery adopt the UDP (User Datagram Protocol) for data transmission and employ FEC (Forward Error Correction) coding as the error-resilient scheme. This work presents a SCTP with different solutions in the per-path status estimation, congestion control, flow rate assignment, and data retransmission.
\begin{table}[htbp]
\small
\setlength{\tabcolsep}{1pt}
\renewcommand{\arraystretch}{1}
      \caption{Basic notations used throughout this paper.}
        \centering
        \renewcommand\arraystretch{1.23}
      \begin{tabular}{|c|l|}
      \hline
       \textbf{Symbol} & \textbf{Definition} \\
        \hline
        \hline
        $\mathbb{P},\mathbb{E}$ & the probability value, expectation value.\\
        \hline
        $\mathcal{P},p$ &  the set of available paths, a path element.\\
        \hline
        $P$ & the number of available paths.\\
        \hline
        $\mathcal{R},R_p$ & the flow rate allocation vector, an element.\\
        \hline
        $\mathcal{T}$, $\Delta$ & the delay, loss requirement.\\
        \hline
        $RTT_{p}$  & the round trip time of $p$.\\
        \hline
        $\mu_{p}$ & the available bandwidth of $p$.\\
        \hline
        $\nu_{p}$ & the residual bandwidth of $p$.\\
        \hline
        $\pi^B_{p}$ & the path loss rate of $p$.\\
        \hline
        $\Pi_{p}^*$ & the effective loss rate over $p$.\\
        \hline
        $\pi_{p}^*$ & the transmission loss rate over $p$.\\
        \hline
        $G/B$ & the Good/Bad state of $p$.\\
        \hline
        $\xi_{p}^G$ &the state transition probability of $p$ from $B$ to $G$.\\
        \hline
        $n,n_p$  & the total number of packets, dispatched onto $p$.\\
        \hline
        $D_{p}$ & the end-to-end delay over path $p$.\\
        \hline
        $\mathcal{U},U_p$ & the system utility matrix, an element.\\
        \hline
      \end{tabular}
      \label{tbl1}
\end{table}
\section{Related Work}
The related work to this paper can be generally classified into two categories: (1) concurrent multipath transfer, and (2) cooperative video delivery in heterogeneous wireless networks. 
\subsection{Concurrent Multipath Transfer}
The authors in [15] generally review the ongoing SCTP standardization progress and gives an
overview of the challenging issues in transport and security. Iyengar et al. [13] investigates the CMT's three negative side effects: 1) unnecessary fast retransmissions; 2) overly conservative congestion window (cwnd) growth; 3) increased acknowledgment traffic. A CMT solution with a potentially failed state (CMT-PF) is proposed by Natarajan et al. [14]. A path that experiences a single timeout is marked as a ``potentially failed'' (PF), indicating the degrees of its communication reliability.

Xu et al. [5] propose a Quality-Aware Adaptive Concurrent Multipath Transfer (CMT-QA) scheme that distributes the data based on estimated path quality. Although the path status is an important factor that affects the scheduling policy, the application requirements should also be considered to guarantee the QoS. Basically, the proposed CMT-DA is different from the CMT-QA as we take the video distortion as the benchmark. Still, the proposed solutions (path status estimation, flow rate allocation, and retransmission control) in CMT-DA are significantly different from those in CMT-QA. In another research conducted by Xu et al. [16], a realistic evaluation tool-set is proposed to analyze and optimize the performance of multimedia distribution when taking advantage of CMT-based multihoming SCTP solutions.

Liao et al. [17] investigate the multipath selection problem in concurrent multipath transfer and propose a path picking scheme to reduce the correlation level. Wang et al. [18] propose a novel out-of-order scheduling approach for in-order arriving of the data chunks in CMT based on the progressive water-filling algorithm.
\subsection{Cooperative Video Delivery in Heterogeneous Wireless Networks}
The existing studies in this field can be divided into: 1) rate allocation policies, e.g., [10][19][20]; 2) packet scheduling approaches, e.g., [8][21][33]; 3) FEC coding schemes, e.g., [7][22].

Zhu et al. [19] study the distributed rate allocation policies for multihomed video streaming over heterogeneous wireless networks and conclude that the media-aware rate allocation policies outperform the heuristic AIMD-based schemes in improving video quality. Freris et al. [10] propose a joint rate control and packet scheduling framework for scalable video streaming from a video server to multihomed clients over heterogeneous wireless networks.

The Earliest Delivery Path First (EDPF) [8] algorithm takes into account the available bandwidth, propagation delay and video frame size for estimating the arrival time and aims to find an earliest path for delivering the video packet. In [21], we propose a novel Sub-Frame Level (SFL) scheduling approach, which deliberately splits the large-size video frames to optimize the delay performance of high definition video streaming.

Han et al. [7] propose an end-to-end virtual path construction system that exploits the path diversity in heterogeneous wireless networks based on fountain code. The Encoded Multipath Streaming [22] model proposed by Chow et al. is a joint multipath and FEC approach for real-time live streaming applications. The authors provide asymptotic analysis and derive closed-form solution for the FEC packets allocation.

In addition to the above researches, Lee et al. [23] study the cost minimization problem and propose a Markov Decision Process (MDP) based solution that includes the parameter estimation, threshold adjustment, and threshold compensation. In [24], a vertical handover decision problem is dealt to minimize an overall communication cost assuming that each of the considered networks can allocate enough resources to the users of interest to prevent buffer underflow.

In conclusion, the existing CMT solutions mainly focus on the performance gains of network-level criteria, e.g., delay, throughput or loss rate. To the best of our knowledge, this is the first work to introduce the video distortion into SCTP for enhancing streaming video quality in heterogeneous wireless environments.
\section{Model and Problem Statement}
As illustrated in Fig. 2, we consider the transmission of a single video flow using the SCTP association form a source node to a multihomed client. The transportation process is activated via invoking the SCTP socket. The encoded video data is divided into several chunks and dispatched onto different paths. The receiver reassembles and reorders the scheduled packets in the receiving buffer to restore the original video traffic for upper-layer applications. The major components at the sender side are the parameter control unit, flow rate allocator, and retransmission controller. The parameter control unit is responsible for processing the ACKs (acknowledgements) feedback from the receiver, estimating the path status and adapting the congestion window size. The delay and loss requirements are imposed by the video applications to achieve the target video quality.

Based on the estimated path status, the rate allocator dynamically picks a subset of communication paths and assigns the transmission rates. The retransmission controller leverages the Explicit Congestion Notification (ECN) [25] to differentiate between network congestion and wireless channel errors. Furthermore, we only retransmit the data chunks which are estimated to arrive at the destination within the deadline.

The flow rate allocation is the critical step in the scheduling procedure. This problem involves the models of communication path and video distortion.
\begin{figure*}[htbp]
\centering
 \includegraphics[width=0.85\textwidth,keepaspectratio]{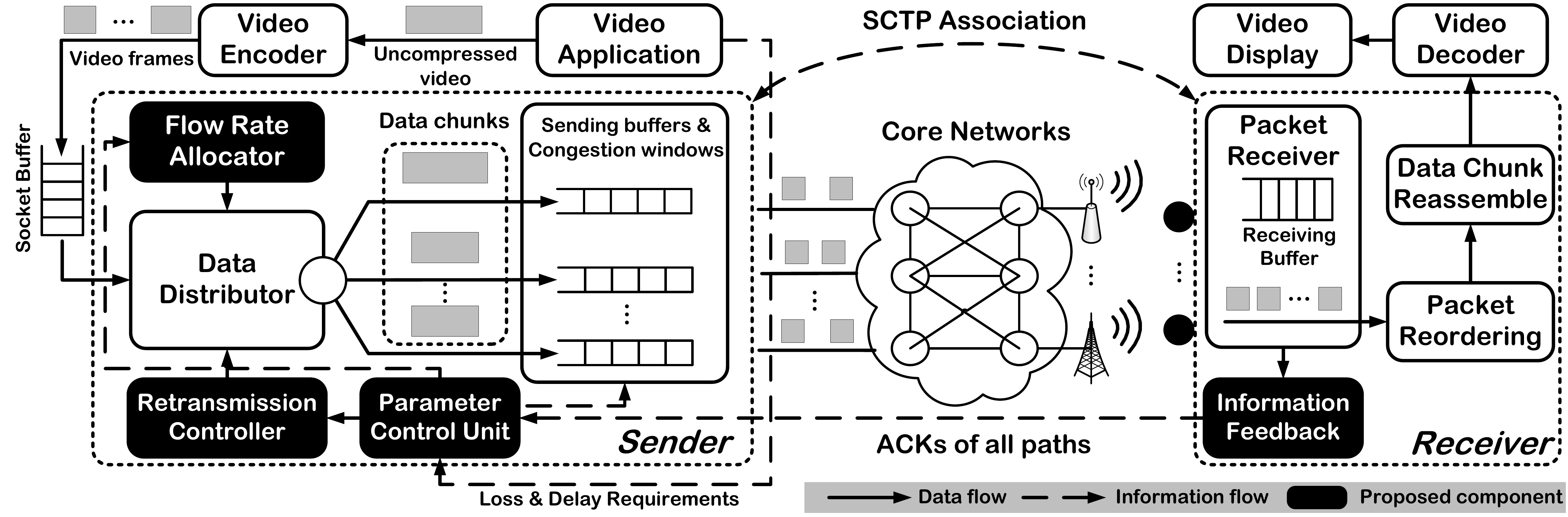}
 \caption{System overview of the proposed CMT-DA solution.}\label{fig2}
\end{figure*}
\subsection{Communication Path Model}
We consider a heterogeneous wireless network integrating $P$ communication paths between two transmission ends. The end-to-end connection can be constructed by binding a pair of IP addresses from the source and destination nodes, respectively. It is well-known that wireless access link is most likely to be the bottleneck for end-to-end transmission due to the limited bandwidth and time-varying channel status. Each communication path $p\in \mathcal{P}$ is considered to be an independent transport link uncorrelated with others and is characterized by the following properties.
\begin{itemize}
  \item The available bandwidth $\mu_p$. This metric does not indicate the raw per-path capacity, but the time-varying share of that bandwidth as perceived by the end-to-end flow. 
  \item The round trip time $RTT_p$, which represents the length of time it takes for a data packet to be sent plus the delay it takes for an acknowledgment of that packet to be received. 
  \item The path loss rate $\pi^B_p$, assumed to be an independent and identically distributed (i.i.d) process, uncorrelated with the input video streaming rate.
\end{itemize}
Similar to the previous work [26], we assume that the background traffic is much larger than our own, and thus the traffic load we impose on a path does not affect its loss statistics.

In addition to the above metrics, we model burst loss on each path using the Gilbert loss model [27], which can be expressed as a two-state stationary continuous time Markov chain. The state $\mathcal{X}_{p}(t)$ at time $t$ assumes one of two values: $G$ (Good) or $B$ (Bad). We assume the path loss rate to be zero in the `Good' state [43][44][45]. If a packet is sent at time $t$ and $\mathcal{X}_{p}(t)=G$ then the packet can be successful delivered. Conversely, if $\mathcal{X}_{p}(t)=B$ then the packet is lost. We denote by $\pi^G_{p}$ and $\pi^B_{p}$ the stationary probabilities that path $p$ is good or bad. Let $\xi_{p}^G$ and $\xi_{p}^B$ represent the transition probability from $B$ to $G$ and $G$ to $B$, respectively. In this work, we adopt two system-dependent parameters to specify the continuous time Markov chain packet loss model: (1) the channel loss rate $\pi^B_p$, and (2) the average loss burst length $1/\xi^B_p$. Then, we will have
\begin{equation}
\begin{split}
\pi_{p}^G = \frac{\xi^G_p}{\xi^B_p+\xi^G_p},\text{ and }
\pi_{p}^B = \frac{\xi^B_p}{\xi^B_p+\xi^G_p}\cdot
\end{split}
\end{equation}
\subsection{Video Distortion Model}
In this subsection, we introduce a generic video distortion model proposed in [28]. The end-to-end distortion ($D_{\text{total}}$) perceived by the end user can be generally computed as the sum of the source distortion ($D_{\text{src}}$) and the channel distortion ($D_{\text{chl}}$). Overall, the end-to-end distortion can thus be written as
\begin{equation}
\begin{split}
D_{\text{total}}=D_{\text{src}}+ D_{\text{chl}}.
\end{split}
\end{equation}
The video quality depends on both the distortion due to a lossy encoding of the media information, and the distortion due to losses experienced in the network. $D_{\text{src}}$ is mostly determined by the video source rate ($R$) and the video sequence parameters (e.g., for the same encoding bit rate, the more complex the sequence, the higher the source distortion). The source distortion decays with increasing encoding rate. The decay is quite steep for low bit rate values, but it becomes very slow at high bit rate. The channel distortion is dependent on the effective loss rate $\Pi$, which is caused by the transmission loss and expired arrivals of video packets. It can be computed as the average of the loss probabilities of all the communication paths. Hence, we can explicitly formulate $D_{\text{total}}$ (expressed in units of Mean Square Error, MSE) as:
\begin{equation}
D_{\text{total}}=D_{0}+\frac{\alpha}{R-R_0}+\beta\cdot\Pi,
\end{equation}
in which $\alpha$, $R_0$, $D_0$, and $\beta$ are constants for a specific video codec and video sequence. These parameters can be estimated from three or more trial encodings using nonlinear regression techniques. To allow fast adaptation of the flow rate allocation to abrupt changes in the video content, these parameters can be updated for each Group of Pictures (GoP) in the encoded video sequence [19]. The encoding parameters (e.g., the frame structure, GoP size, etc.) also affect the source and channel distortion. But they are not used as control parameters as the proposed CMT-DA is a transport-layer protocol. Since this model takes into account the effects of intra coding and spatial loop filtering, it provides accurate estimations for end-to-end distortion [28][48].

Recent studies [10][49] propose to model the channel distortion of each video frame with the truncation and channel distortion. This model is able to estimate the distortion caused by the inter-frame dependency. However, it also needs the offline estimation of the distortion parameters.

Finally, we should note that the perceived video quality is not just correlated with the effective loss rate. As future work, we will consider the contexts (e.g., user mobility, viewing distance, video motion degree, etc.) to optimize the Quality of Experience (QoE) perceived by end users.
\subsection{Derivation of Effective Loss Rate}
The effective loss rate represents the combined probability of transmission and overdue losses. Therefore, the effective loss rate over path $p$ can be expressed as follows
\begin{equation}
\Pi_p=\pi_{p}^*+(1-\pi_{p}^*)\cdot \mathbb{P}\{D_p>\mathcal{T}\}.
\end{equation}
First, we provide derivations of $\pi_p^*$ based on continuous time Markov chain and Gilbert loss model.
\subsubsection{Transmission Loss Rate}
In the scheduling process of SCTP, the encoded video data (e.g., a Group of Pictures with the total size $S$) is divided into multiple chunks ($S_p, p\in \mathcal{P}$) and each of them is dispatched onto a different path. The chunks will be fragmented into packets when transmitted over each communication path. Therefore, the number of packets on each path can be estimated with
\begin{equation}
n_p=\left\lceil\frac{S_p}{MTU}\right\rceil,
\end{equation}
in which $\lceil x\rceil$ represents the smallest integer larger than $x$ and $MTU$ denotes the Maximum Transmission Unit size. The path MTU probing and management approach can be referred to RFC 5061 [29]. We assume the packets over each path are evenly spread with interval $\omega_p$. Let $c$ denote a $n$-tuple that represents a specific path failure configurations. If the $i$th packet dispatched onto path $p$ is lost, then $c_p^i=B, p\in \mathcal{P},1\leq i\leq n_p$ and vice versa. By taking into account all the possible configurations of $c_p$, we can obtain the transmission loss rate $\pi_p^*$ as
\begin{equation}
\pi_p^*=\frac{1}{n_p}\cdot\sum_{\text{all } c_p}L(c_p)\cdot \mathbb{P}(c_p),
\end{equation}
in which ``all $c_p$'' denotes all the possible combinations of $c_p$ and $L(c_p)$ represents the number of lost packets on path $p$. $L(c_p)$ can be expressed as
\begin{equation}
L(c_p)=\sum_{i=1}^{n_p}1_{\{c_p^i=B\}}.
\end{equation}
Suppose $\mathbb{P}(c_p^i)$ denotes the probability of path failure on $p$ when delivering the $i$th packet. We can derive $\mathbb{P}(c_p^i)$ based on continuous time Markov chain and Gilbert loss model. Let $f_p^{i,j}(\omega_p)$ denote the probability of transition from state $i$ to $j$ on path $p$ in time $\omega_p$
\begin{equation}
f_p^{i,j}(\omega_p)=\mathbb{P}[\mathcal{X}_p(\omega_p)=j|\mathcal{X}_p(0)=i].
\end{equation}
According to the transient behaviour of continuous time Markov chain, the state transition matrix can be expressed as follows
\begin{equation}
\begin{split}
&f_{p}^{G,G}(\omega_p)=\pi^G_{p}+\pi^B_{p}\cdot\kappa_p ,\,\, f_{p}^{G,B}(\omega_p)=\pi^B_{p}-\pi^B_{p}\cdot\kappa_p,\\
&f_{p}^{B,G}(\omega_p)=\pi^G_{p}-\pi^G_{p}\cdot \kappa_p,\,\,
f_{p}^{B,B}(\omega_p)=\pi^B_{p}+\pi^G_{p}\cdot \kappa_p,
\end{split}
\end{equation}
where $\kappa_p=\exp\left[-\left(\xi_p^B+\xi_p^G\right)\cdot \omega_p\right]$. Now, we can have the expression of $\mathbb{P}(c_p)$ as
\begin{equation}
\mathbb{P}(c_p)=\prod_{i=1}^{n_p}\mathbb{P}(c_p^i)=\pi_p^{c_p^i}\cdot\prod_{i=1}^{n_p-1}\left(f_{p}^{c_p^i,c_p^{i+1}}(\omega_p)\right).
\end{equation}
Finally, $\pi_p^*$ can be obtained with
\begin{equation}
\pi_p^*=\left\lceil\frac{MTU}{S_p}\right\rceil\cdot\sum_{\text{all }c_p}\pi_p^{c_p^i}\cdot\prod_{i=1}^{\left\lceil\frac{S_p}{MTU}\right\rceil-1}\left(f_{p}^{c_p^i,c_p^{i+1}}(\omega_p)\right).
\end{equation}
\subsubsection{Overdue Loss Rate}
The overdue loss rate represents the ratio of packets arriving at the destination out of the deadline and is determined by the end-to-end delay. The end-to-end transmission delay ($D_p$) is dominated by the queueing delay at the bottleneck link and can be approximated by an exponential distribution [19][30], i.e.,
\begin{equation}
\mathbb{P}\left\{D_p>\mathcal{T}\right\}=\exp\{-\Theta\cdot \mathcal{T}\},
\end{equation}
where $\Theta$ represents the arriving rate and is inversely proportional to the average delay, i.e.,
\begin{equation}
\Theta=\frac{1}{\mathbb{E}\{D_p\}},
\end{equation}
where $\mathbb{E}\{\cdot\}$ represents the expectation value. Generally, $\Theta$ needs to be empirically determined from end-to-end delay statistics. In order to derive a general solution for online operation, we construct a model to approximate the average packet delay. We suppose the flow rate assignment to be expressed in a vector form, i.e., $\mathcal{R}=\{R_p\}_{p\in \mathcal{P}}$, in which element $R_p$ represents the assigned flow rate over communication path $p$. We denote the residual bandwidth of $p$ with $\nu_p$. Then, we can have
\begin{equation}
\nu_p=\mu_p-R_p.
\end{equation}
As the assigned video streaming rate on each path approaches the available bandwidth, the average packet delay typically increases due to network congestion. We use a fractional function to approximate the delay of the allocated sub-flow rate $R_p$ over path $p$, i.e.,
\begin{equation}
\mathbb{E}\left\{D_p\right\}=\frac{R_p}{\mu_p}+\frac{\rho_p}{\nu_p},
\end{equation}
in which $\rho_p$ can be interpreted as the available source of communication path $p$. The value of $\rho_p$ can be estimated from the latest observations of the path status information
\begin{equation}
\rho_p=\frac{\nu'_p\cdot RTT_p}{2}.
\end{equation}
If $\nu'_p$ is equal to the latest observed residual bandwidth of path $p$, i.e., $\nu'_p=\nu_p$, the one-way delay is $RTT_p/2$. Then, we can have
\begin{equation}
\begin{split}
\mathbb{P}\left\{D_p>\mathcal{T}\right\}=\exp\left\{-\frac{2\times \mathcal{T}\cdot \nu_p \cdot\mu_p}{\nu'_p\cdot RTT_p\cdot\mu_p+2\times \nu_p\cdot R_p}\right\}.
\end{split}
\end{equation}

\subsection{Problem Formulation}
The expected effective loss rate of all the paths can be estimated with $\sum_{p\in \mathcal{P}}(R_p\cdot\Pi_p)/\sum_{p\in \mathcal{P}}R_p$ and the channel distortion can be expressed as
\begin{equation}
D_{\text{chl}}=\beta\cdot\frac{\sum_{p\in \mathcal{P}}R_p\cdot\left\{\pi_p^*+\left(1-\pi_p^*\right)\cdot\mathbb{P}\left\{D_p>\mathcal{T}\right\}\right\}}{\sum_{p\in \mathcal{P}}R_p}.
\end{equation}
In conjunction with the source distortion (Equation (3)), we can obtain the end-to-end video distortion as follows
\begin{equation}
D_{\text{total}}(\mathcal{R})=D_0+\frac{\alpha}{R-R_o}+\beta\cdot\frac{\sum_{p\in \mathcal{P}}R_p\cdot\Pi_p}{\sum_{p\in \mathcal{P}}R_p}.
\end{equation}
To reduce the probability of out-of-order packet arrivals, the scheduling policy aims at minimizing the delay jitters of different paths. We should note that the video encoding rate $R$ is not a fixed parameter for CBR (Constant Bit-Rate) video streaming due to the bit rate variability [46][47]. Therefore, the actual bit rate is dynamically estimated with the division of input data size to distribution interval. Now, we are ready to formulate the constrained optimization problem for given communication paths $\mathcal{P}$ on minimizing the end-to-end distortion of input video streaming while satisfying the path capacity, delay constraint, etc.
\begin{displaymath}
\begin{split}
&\textbf{\text{For each data distribution interval,}} \\
&\textbf{\text{determine the values of }}\mathcal{R}=\left\{R_p\right\}_{p\in \mathcal{P}}\\
&\textbf{\text{to minimize: }}D_{\text{total}}=D_0+\frac{\alpha}{R-R_o}+\beta\cdot\frac{\sum_{p\in \mathcal{P}}R_p\cdot\Pi_p}{\sum_{p\in \mathcal{P}}R_p},
\end{split}
\end{displaymath}
\begin{equation}
\text{s. t.}
\begin{cases}
\Pi_{p}=\pi^*_p+(1-\pi^*_p)\cdot\mathbb{P}\left\{D_p>\mathcal{T}\right\}, p\in \mathcal{P},\\
\pi^*_p=\text{Equation (11)},\\
\mathbb{P}\left\{D_p>\mathcal{T}\right\}=\text{Equation (17)},\\
\mathbb{E}\left\{D_p\right\}+\left(\left\lceil\frac{S_p}{MTU}\right\rceil-1\right)\cdot \omega_p\leq \mathcal{T}, p\in \mathcal{P},\\
R_p\leq \mu_p, p\in \mathcal{P},\\
\mathbb{E}\left\{D_p\right\}=\mathbb{E}\left\{D_{p'}\right\}, p'\neq p,\{p,p'\}\in \mathcal{P}.
\end{cases}
\end{equation}
To obtain the close-to-optimal result with fast convergence for efficient online operation, we propose a progressive flow rate allocation algorithm to solve the optimization problem based on the utility maximization theory [31]. In the next section, we will describe the solution procedure in detail.

\section{Proposed CMT-DA Solution}
\subsection{Path Status Estimation and Congestion Control}
As introduced in Section 3.2, the communication path model includes the status of available bandwidth, round trip time, and path loss rate. The path loss rates are updated once an ACK is received on any path. It represents the ratio of successfully delivered packets at the destination to the total number of packets dispatched ont path $p$. The state transition matrix can be estimated following the Table 3 in [32]. The per-path $RTT_p$ is estimated the same way as performed in TCP. Then, the available bandwidth can be estimated as $\mu_p=cwnd_p/RTT_p$.

In the environment of heterogeneous access networks, the different physical characteristics or network conditions will result in path asymmetry. The uniform congestion control may induce performance degradation caused by the paths with lower quality. Consequently, the congestion control is independently performed for each path, uncorrelated with other paths.

In the design of CMT-DA, each communication path employs the following congestion control parameters: congestion window size ($cwnd_p$), round trip time ($RTT_p$), and retransmission timeout ($RTO_p$). The ACKs in CMT-DA are provided on the aggregate level. The latest aggregate feedback is sent by the destination upon the receipt of each packet. An interesting feature in the per-path congestion control is that it allows the ACK packets to be sent back through any uplink path, i.e., not necessarily the path on which the last packet is received by the destination. The receiver sends the ACK on a most reliable uplink path and this reduces the probability of dropped/overdue feedback packets. The purpose for picking the most reliable uplink channel is also to reduce the  round trip time.

For a specific communication path, the aggregate selective/cumulative ACK feedback is filtered to obtain the delivery status of the scheduled packets. The per-path ACK information is used for sliding the congestion window of the path. CMT-DA adjusts the value of $cwnd_p$ based on the ECN. Note that the window size adaption strategy is different from that of TCP as it is based on the value of packet acceptance ratio, i.e., $1-\pi_p$.

A path congestion controller will timeout if no feedback is received for a period $RTO_p$. Therefore, the timeouts are handled separately for each path. The response to a timeout is identical to conventional TCP response to a timeout. CMT-DA does not perform fast retransmission after a certain number of dup-ACKs are received. Fast retransmission in response to dup-ACKs could result in a lot of unnecessary transmissions. The fast retransmission can be improved to be more efficient if SACK (selective acknowledgement) information is taken into account.

\subsection{Flow Rate Allocation}
In this subsection, we firstly employ the piecewise linear (PWL) approximation to derive the potential rate allocation vectors to minimize the end-to-end video distortion. Then, we use the utility based rate allocation function to minimize the video distortion while alleviating load imbalance.

As $D_{\text{total}}$ is dependent on the sum of the transmission and overdue loss rate over each communication path, a PWL approximation can be obtained based on a univariate function. It can be achieved by dividing the interest region of each univariate function into a sufficient number of non-overlapped small intervals. Let $l(\cdot)$ denote an univariate function with the interest region of $[a,a']\subset \mathbb{R}$. We assume $m$ breakpoints in the region are appropriately chosen so that $l(\cdot)$ can be approached by the function $\hat{l}(k)=A_k\cdot x+B_k$ in each small interval $I_k=[a_{k-1},a_k],\,1\leq k\leq m+1$. $A_k$ and $B_k$ are determined by the linear equations $l(a_{k-1})=\hat{l}(a_{k-1})$ and $l(a_{k})=\hat{l}(a_{k})$. The approximation function $\varphi(\cdot)$ of $l(\cdot)$ on $[a,a']$ can be obtained via connecting these intervals.

For any $1\leq k \leq m$, we name $a_k$ an turning point if $A_k>A_{k+1}$. Let $a_{t(1)}<\ldots<a_{t(q)}$ denote all the turning points among the breakpoints $[a_1,a_m]$. We define $\hat{I}_t=[a_{t(i-1)},a_{t(i)}], a\leq i\leq q+1$. It can be observed that $\hat{I}_t$ is the union of intervals $I_k,t(i-1)<k<t(i)$. Based on the aforementioned partition, we can obtain a piecewise-convex expression of the function $\varphi(\cdot)$, which is very useful to obtain the global optimization of separable programming problems. Then, we can have
\begin{equation}
\varphi(\lambda)=\max\left\{\hat{l}_k(\lambda)\right\},\forall \lambda \in \hat{I}_t, t(i-1)<k<t(i).
\end{equation}

Since $A_{t(i-1)+1}<A_{t(i-1)+2}<\ldots<A_{t(i)}$, the function $\varphi(\cdot)$ is convex on $\hat{I}_t$ for any $\hat{\lambda}\in \hat{I}_t$. We can choose $\bar{\lambda}\in  \hat{I}_{k'}\subset \hat{I}_{t}$ and a small positive number $\epsilon$ such that
\begin{equation}
\begin{split}
\varphi(\lambda')&\leq \epsilon\cdot \varphi(\hat{\lambda})+(1-\epsilon)\cdot\varphi(\bar{\lambda}),\\
&=\epsilon\cdot \hat{l}_k(\hat{\lambda})+(1-\epsilon)\cdot\hat{l}_k(\bar{\lambda}),\\
&<\epsilon\cdot \hat{l}_{k'}(\hat{\lambda})+(1-\epsilon)\cdot\hat{l}_{k'}(\bar{\lambda})=\hat{l}_{\lambda'}.
\end{split}
\end{equation}

Then, we can approximate the goal function on every hypercube by a convex PWL function. In fact, the goal function $D_{\text{total}}$ corresponds to the arbitrary univariate function $l(\cdot)$ and any potential rate allocation vector is the breakpoint of the PWL function. Therefore, how to find the appropriate breakpoints and judge whether it is an inflection point is the key point for implementing the piecewise approximation of the video rate allocation problem. We seek to solve the difficulty by employing a utility-based function in the rate allocation algorithm.

To minimize the end-to-end video distortion, the rate allocation algorithm is inclined to assign loads to the communication paths with higher quality. The quality level offered by a path is proportional to its transmission ability. However, this policy will in turn result in load imbalance and link congestion during the transmission process. To alleviate severe load imbalance problems, we introduce a load imbalance parameter $L_p$ to indicate whether path $p$ is overloaded and it can be expressed as
\begin{equation}
L_p=\frac{\mu_p\cdot(1-\pi_p)-R_p}{\left(\sum_{p\in \mathcal{P}}\mu_p\cdot(1-\pi_p)-\sum_{p\in \mathcal{P}}R_p\right)/P},
\end{equation}
in which $\mu_p\cdot(1-\pi_p)$ denotes the `loss-free' bandwidth of $p$. When the value of $L_p$ is obviously higher than a threshold limit value (TLV) [19], path $p$ is overloaded. We assume the initial rate allocation for each path is proportional to the available bandwidth, i.e., $R_p=R\cdot \mu_p/\sum_{p\in \mathcal{P}}\mu_p$. Let $\Delta R_p$ denote the rate variation over path $p$ at each iteration and $R_p+\Delta R_p$ represent the transition of the next allocation. The utility of this transition can be expressed as [32]:
\begin{equation}
U_p=\frac{\varphi\left(R_p+\Delta R_p\right)-\varphi(R_p)}{\Delta R_p},
\end{equation}
in which $\varphi(\cdot)$ represents the approximate linear function for $D_{\text{total}}$ in the interval $[R_p, R_p+\Delta R_p]$. The utility matrix can be expressed as $\mathcal{U}=\{U_p\}_{p\in \mathcal{P}}$. In each iteration, the proposed flow rate assignment algorithm obtains the $R_p^f$ that brings the highest utility, i.e.,
\begin{equation}
\overline{\mathcal{U}}=\underset{\mathcal{R}}{\operatorname{argmax}}\{\mathcal{U}\}.
\end{equation}
The proposed algorithm allocates the channel resources available in heterogeneous wireless networks in an iterative manner. Once the resources of path $p$ are exhausted, the algorithm will seek a different path which can release the required resources for the video flow. This operation will be performed until utility value of the system can not be improved or the the channel resources available are depleted. The sketch of the iterative flow rate allocation algorithm based utility maximization theory is presented in Algorithm 1.
\begin{algorithm}[htbp]
\small
\caption{\small Utility maximization based flow rate allocation}
\begin{algorithmic}[1]
\REQUIRE $\{RTT_p,\mu_p,\pi_p\}_{p\in \mathcal{P}}$, $\mathcal{T}$, $\Delta$;\\
\ENSURE $\mathcal{R}=\{R_p\}_{p\in \mathcal{P}}$;\\
\FOR{each path $p$ in $\mathcal{P}$}
\STATE $U_p\Leftarrow\frac{\varphi(R_p+\Delta R_p)-\varphi(R_p)}{\Delta R_p}$;
\STATE $L_p\Leftarrow\frac{\mu_p\cdot\left(1-\pi_p\right)-\sum_{f\in \mathcal{F}}R_p}{\left(\sum_{p\in \mathcal{P}}\mu_p\cdot\left(1-\pi_p\right)-\sum_{p\in \mathcal{P}}R_p\right)/P}$;
\STATE $\Delta R_p\Leftarrow \Delta R_p/U_p$;
\STATE $R_p\Leftarrow R_p+\Delta R_p$;
\STATE Update the approximate function $\varphi(R_p)$;
\ENDFOR
\STATE $\overline{\mathcal{U}}\Leftarrow\underset{\mathcal{R}}{\arg\max}\{\mathcal{U}\}$;
\IF{$L_p\leq$TLV}
\STATE $R_p\Leftarrow R_p+\Delta R_p$ $//$Intra-path allocation;
\STATE Update free resources of path $p$;
\ELSE
\STATE find other path that can transfer part of its assigned rate to path $p'\neq p\in \mathcal{P}$ with maximum transition utility improvement $\Delta \mathcal{U}$ $//$Inter-path allocation;
\IF{ $\Delta \mathcal{U}>0$ }
\STATE $R_p\Leftarrow R_p+\Delta R_p$;
\STATE Update free resources of path $p$ and $p'$;
\ENDIF
\ENDIF
\end{algorithmic}
\end{algorithm}

In this algorithm, the intra-path allocation process always attempts to increase the system¡¯s utility by assigning some resources in path $p$. If the available channel resources are not adequate, this procedure will find a new path that can allocate parts of its unassigned resources to the current path. After the flow rate allocation vector is determined, we can easily enforce them on the chunk size for each communication path, i.e., $S_p=S\cdot R_p/\sum_{p'\in \mathcal{P}}R_{p'}$.


During the flow rate allocation interval, when the first data chunk is to be sent, the sending time is recorded as the starting time for a successful transmission. If a data chunk is sent to any path (including a retransmission) and the path congestion controller has not been started yet, the sender starts it immediately so that it expires after the $RTO$ of that path. If the timer for that path is already running, the sender restarts the timer whenever an outstanding data chunk previously scheduled onto that path is retransmitted. Whenever a SACK is received that acknowledges the data chunk with an outstanding transmission sequence number (TSN) for that path, the path congestion controller is restarted for that path with its current RTO (if there is still outstanding data on that path). If all outstanding data sent to a path has been acknowledged, the path congestion controller should be turned off for that path. If packet loss occurs, either detected by the retransmission timeout or reported as missing by duplicate SACKs, the sender should retransmit the lost chunk immediately, to mitigate the reordering. In the packet loss case, the timestamp (indicating the sending time) of the last data chunk is recorded as the end time of the successful data transmission interval.
\subsection{Data Retransmission Control}
The data retransmission is necessary to satisfy the imposed loss requirements for achieving acceptable video quality and SCTP standard defines two retransmission schemes: fast retransmission and timeout retransmission. The retransmission process is activated when packet loss occurs in one path, either detected by the SACKs on gap report or after a retransmission timeout without acknowledgment.

The SCTP sender uses a path congestion controller to guarantee data delivery in the absence of any feedback from the receiver side. If the controller of communication path $p$ expires, the value of $cwnd_p$ is set to be the MTU size and the end host enters the slow start mode. A retransmission timeout will double the $RTO_p$, whereas a successful retransmission will not refresh the $RTO_p$ which can only be updated by the heartbeat chunks. Consequently, the $RTO_p$ is usually a large value which causes the data loss detection time to become very long and degrades the delivery performance. By introducing a fast retransmission function, loss can be recovered rapidly and the delivery quality for the users can be maintained at high levels. Fast retransmission helps to avoid the long waiting time for the retransmission timer to expire and reduces the average delay. Fast retransmission is considered if SACK indicates that a segment has been missing four times and therefore packet loss has occurred.

When packet losses occur in the condition of concurrent multipath transfer, it will reduce the transmission efficiency of the current path through sharply decreasing the $cwnd_p$. Meanwhile, the existing mechanism does not make a distinction between random packet loss in wireless networks and the congestion loss. The long period to detect the timeout packet in path failures also decreases the transmission efficiency. In conclusion, there is a definite need to design new strategies to handle packet loss more efficiently.

The standard CMT retransmission policy has no mechanism to distinguish random losses from congestion, and therefore treats all losses as congestion induced. In the context of heterogeneous wireless networks, the packet losses can be classified into three categories: 1) congestion loss caused by bandwidth limitations or router buffer overflows; 2) packet errors due to external interference or wireless channel fading; and 3) link failure or handover loss. In wireless networks, most packet losses are caused by the bit errors due to channel fading and not due to congestion [5][7]. 

The delay constraint is seldom considered in exiting SCTP retransmission schemes. Motivated by optimizing the SCTP retransmission policy for real-time video applications, this paper proposes a data retransmission policy under the delay and loss requirements imposed by upper-layer video applications. The data loss rate under $1\%$ is acceptable for most live video applications [22]. On one hand, reducing the number of unnecessary retransmissions can minimize the bandwidth consumption; On the other hand, it can also contribute to alleviate the network congestion and reduce the channel distortion. If timeout occurs on a path, it should be marked with an inactive status. To reduce the detection latency of link status change, a periodic heartbeat packet is sent to check whether links are alive or not. Meanwhile in both cases, the sender tries to reach the active path with the minimum value of the transfer delay to transmit these lost packets as soon as possible. The detailed descriptions of the proposed retransmission policy are summarized in Algorithm 2.
\begin{algorithm}[htbp]
\small
\caption{\small Delay and loss controlled data retransmission}
\begin{algorithmic}[1]
\IF {$ECN_p==true$}
\IF {the congestion controller expires after $RTO_p$}
\STATE $ssthresh_p =\max (cwnd_p/2, 3\times MTU)$;
\STATE $cwnd_p=MTU$;
\ENDIF
\IF {received $3$ dup-SACKs}
\STATE $ssthresh_p =\max (cwnd_p/2, 3\times MTU)$;
\STATE $cwnd_p=ssthresh_p$;
\ENDIF
\IF {recorded $\Pi'>\Delta$}
\STATE $p'=\arg\min\limits_{p\in \mathcal{P}}\left\{\mathbb{E}\{D_{p}(S_p)\}\right\}$;
\IF {$E\left\{D_{p'}(S_p)\right\}<\mathcal{T}$}
\STATE Retransmit the lost data chunk through $p'$;
\ENDIF
\ENDIF
\ENDIF
\end{algorithmic}
\end{algorithm}
\section{Performance Evaluation}
In this section, we firstly describe the evaluation methodology that includes the emulation setup, performance metrics, reference schemes, and emulation scenario. Then, we depict and discuss the evaluation results in detail.
\subsection{Evaluation Methodology}
\subsubsection{Emulation Setup}
We adopt the Exata and JSVM as the network emulator and video codec, respectively. The architecture of evaluation system is presented in Fig. 3 and the main configurations are set as follows.
\begin{figure}[htbp]
\centering
 \includegraphics[width=0.45\textwidth,keepaspectratio]{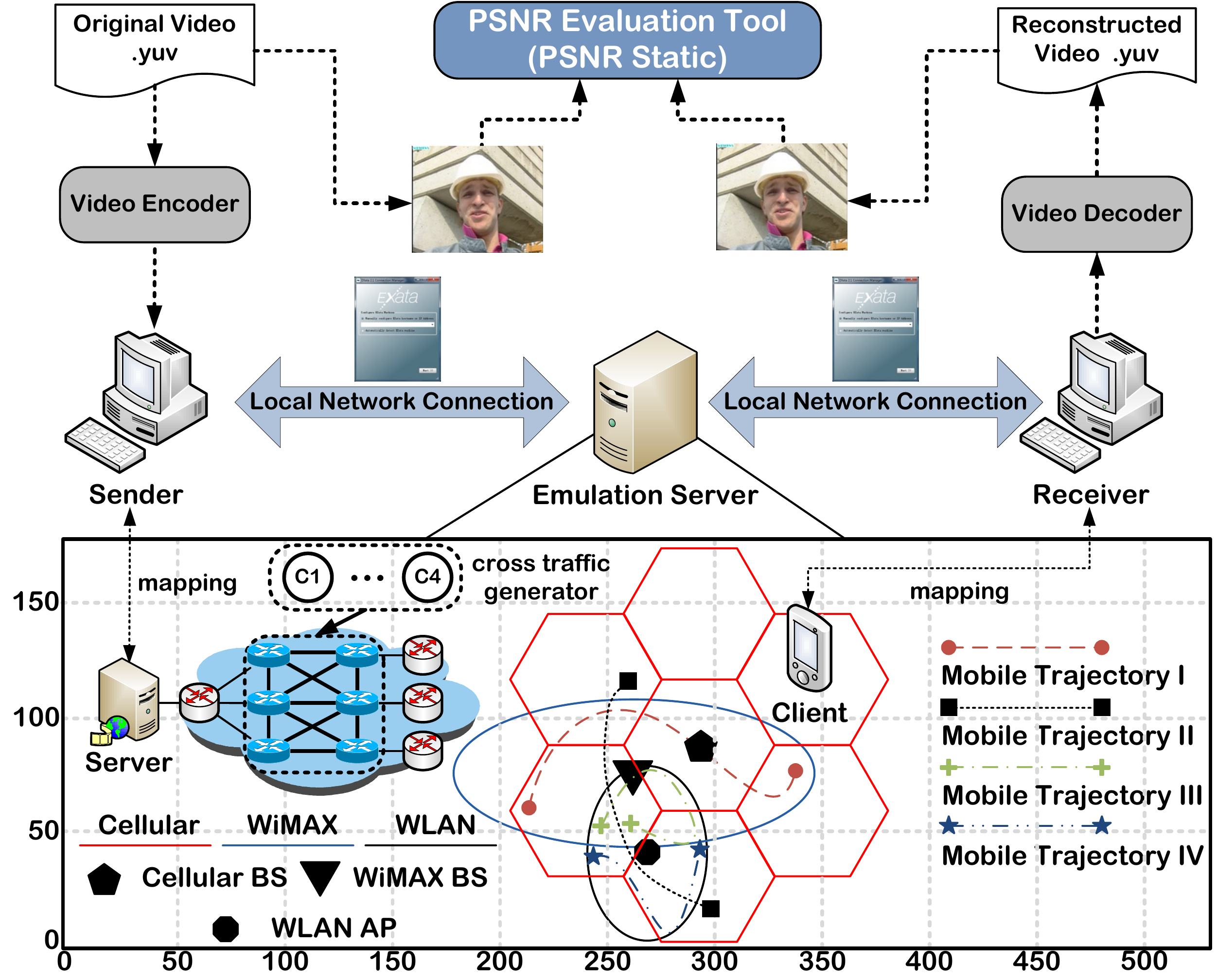}
 \caption{System architecture for performance evaluation.}\label{fig1}
 \end{figure}

\underline{Network emulator}. Exata 2.1 [34] is used as the network emulator. Exata is an advanced edition of QualNet [35] in which we can perform semi-physical emulations. In order to implement the real-video-streaming based emulations, we integrate the source code of JSVM\footnote{We choose the JSVM in convenience for the source code integration as both Exata and JSVM are developed using the C++ code while the H.264/AVC JM [36] software is developed using C language.} [as \emph{Objective File Library} (.LIB)] with Exata and develop an application layer protocol of ``Video Transmission''. The detailed descriptions of the development steps could be referred to Exata Programmer's Guide [34].

In the emulated network topology, the sender has one wired network interface and the mobile client has three wireless network interfaces, i.e., Cellular, WLAN and WiMAX. The parameter configurations of different wireless networks are summarized in Table 2 [37][38]. As the Exata emulator does not include SCTP in the transport-layer stacks, we modify the QualNet simulation module developed by University of Delaware [39] to implement the CMT solutions. As depicted in the figure, each router is attached to one edge node, which is single-homed and introduces background traffic. Each of the edge nodes has four traffic generators producing cross traffic with a Pareto distribution. The packet sizes of background traffic are varied to mimic the real traces collected on the Internet: $50\%$ of them are $44$-Byte long, $25\%$ have $576$ Bytes, and $25\%$ are $1500$-Byte long [40]. The aggregate cross traffic loads imposed on the available network paths are similar and vary randomly between $0-10$ percent of the bottleneck links' bandwidth. The data distribution interval is $250$ ms (the duration of a GoP) and the TLV is $1.2$ [19]. The packet interleaving level ($\omega_p$) is $5$ ms for each path.

\begin{table}[htbp]
        \scriptsize
        \renewcommand{\arraystretch}{1}
      \caption{Parameter configurations of wireless networks}
        \centering
        \begin{tabular}{|c||c|}
        \hline
        \textbf{Cellular parameter} & \textbf{Value}\\
        \hline
        \hline
        Target SIR value & $10$ dB\\
        \hline
        Orthogonality factor & $0.4$\\
        \hline
        Common control channel power & $33$ dB\\
        \hline
         Maximum power of BS & $43$ dB\\
        \hline
         Total cell bandwidth & $3.84$ Mc/s\\
        \hline
         Inter/intra cell interference ratio & $0.55$\\
        \hline
         Background noise power & -$106$ dB\\
         \hline
        available capacity & $300$ Kbps\\
        \hline
         average loss rate & $2\%$\\
         \hline
         average burst length & $10$ ms\\
        \hline
        \hline
        \textbf{WiMAX parameter} & \textbf{Value}\\
        \hline
        \hline
        System bandwidth & $7$ MHz\\
        \hline
        Number of carriers & $256$\\
        \hline
        Sampling factor & $8/7$\\
        \hline
        Average SNR & $15$ dB\\
        \hline
        Symbol duration & $2048$\\
        \hline
        available capacity & $1200$ Kbps\\
         \hline
         average loss rate & $4\%$\\
         \hline
        average burst length & $15$ ms\\
        \hline
        \hline
        \textbf{WLAN parameter} & \textbf{Value}\\
        \hline
        \hline
        Average channel bit rate & $2$ Mbps\\
        \hline
        Slot time & $10$ $\mu$s\\
        \hline
        Maximum contention window & $32$\\
        \hline
        available capacity & $500$ Kbps\\
        \hline
         average loss rate & $6\%$\\
         \hline
         average burst length & $20$ ms\\
         \hline
      \end{tabular}%
 \end{table}

\underline{Video codec}. H.264/SVC reference software JSVM 9.18 [41] is adopted as the video encoder. The generated video streaming is encoded at $30$ frames per second and a GoP consists of $8$ frames. The GoP structure is IPPP. The test video sequences are QCIF (Quarter Common Interchange Format) \emph{Foreman, Bus, Stefan, and Soccer} of $300$ frames-long. Each of the sequences features a different pattern of temporal motion and spatial characteristics that reflected in their corresponding video quality versus encoding rate dependencies. For instance, the \emph{Foreman} sequence has moderate movement and video texture while the \emph{Stefan} has fast movement with different motion directions. We concatenate them $10$ times to be $3000$ frame-long in order to obtain statistically meaningful results. The loss requirement ($\Delta$) and delay constraint ($\mathcal{T}$) are set to be $1\%$ and $250$ ms [7][22], respectively. The streaming video is encoded at the source rates of $1.4$, $1.2$, $1.8$, and $0.65$ Mbps for the Trajectory I to IV. The available capacities are just enough or very tight for delivering the encoded video streaming.
\begin{figure*}[htbp]
\centering
\begin{minipage}[t]{0.5\linewidth}
\centering
 \includegraphics[width=1\textwidth,keepaspectratio]{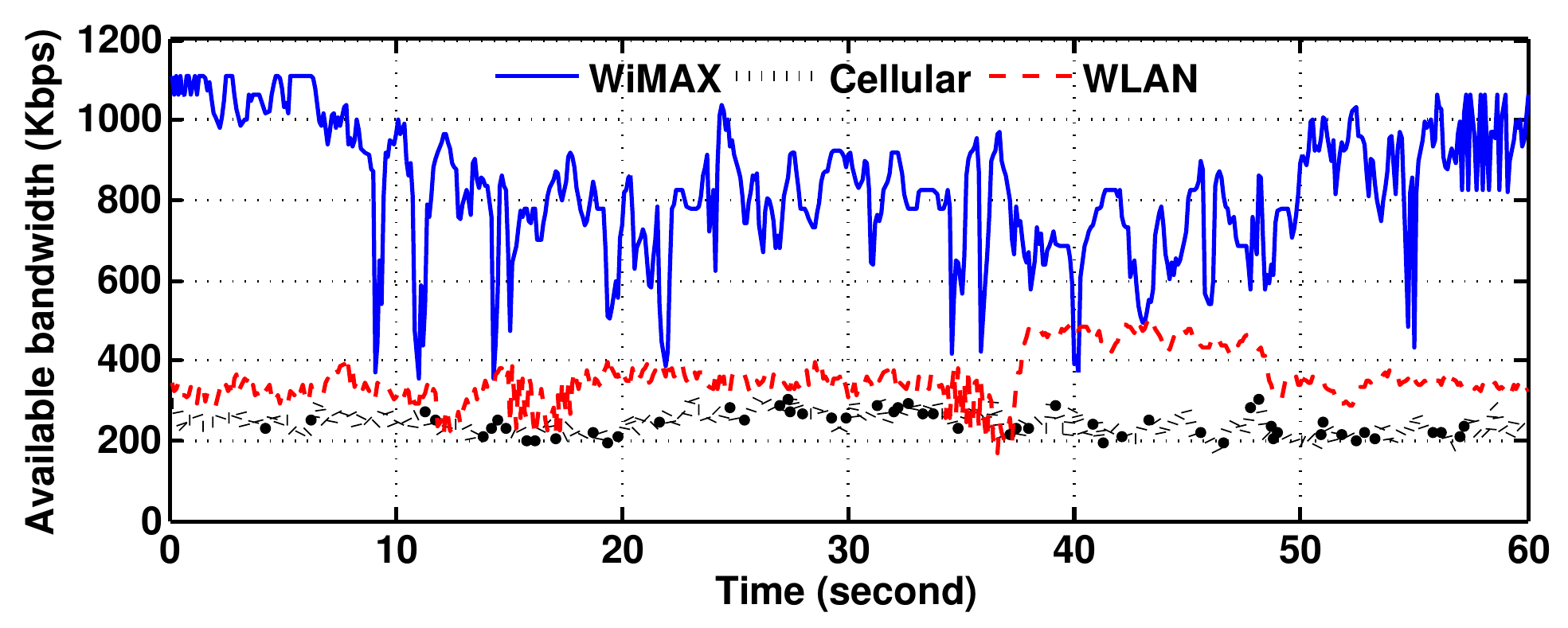}
  \label{4a}
 \end{minipage}%
\begin{minipage}[t]{0.5\linewidth}
\centering
 \includegraphics[width=1\textwidth,keepaspectratio]{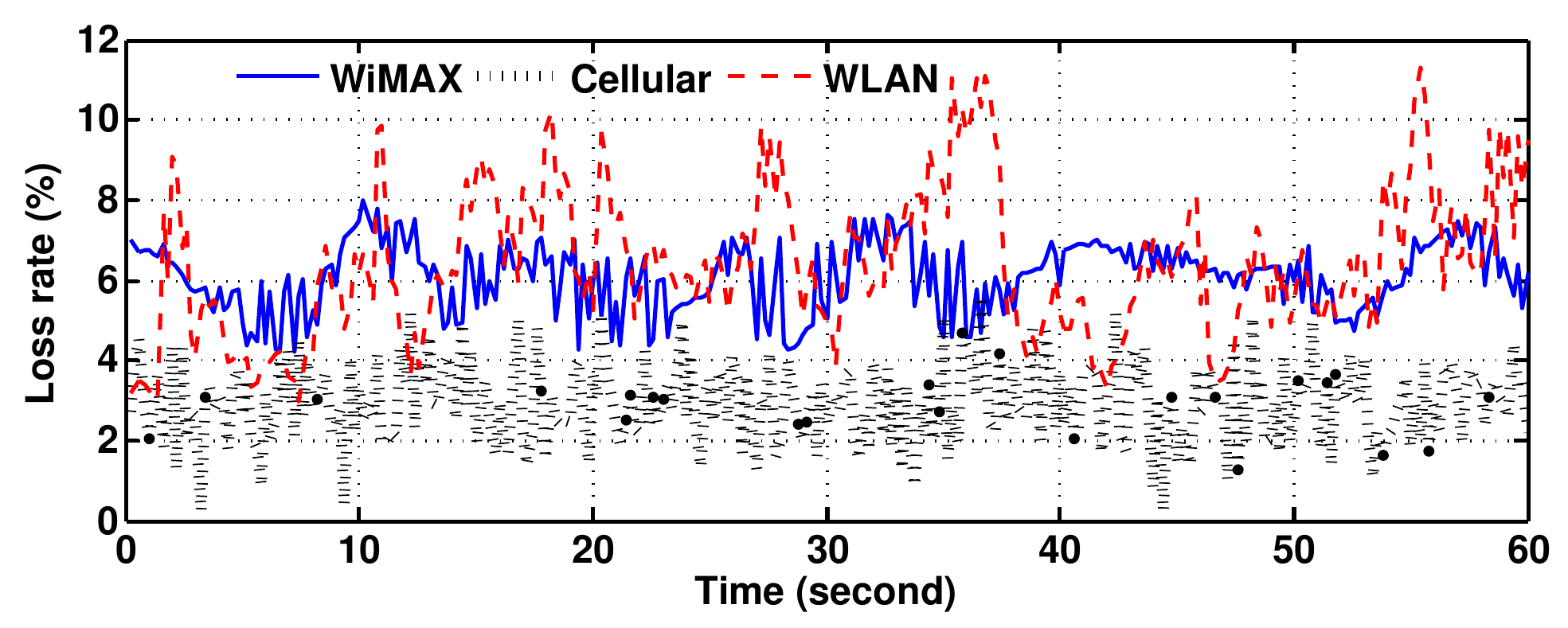}
 \label{4b}
\end{minipage}%
\caption{Profile of channel status information: (a) available bandwidth, (b) path loss rate.}
\end{figure*}
\begin{figure*}[htbp]
\centering
\begin{minipage}[t]{0.25\linewidth}
\centering
 \includegraphics[width=1\textwidth,keepaspectratio]{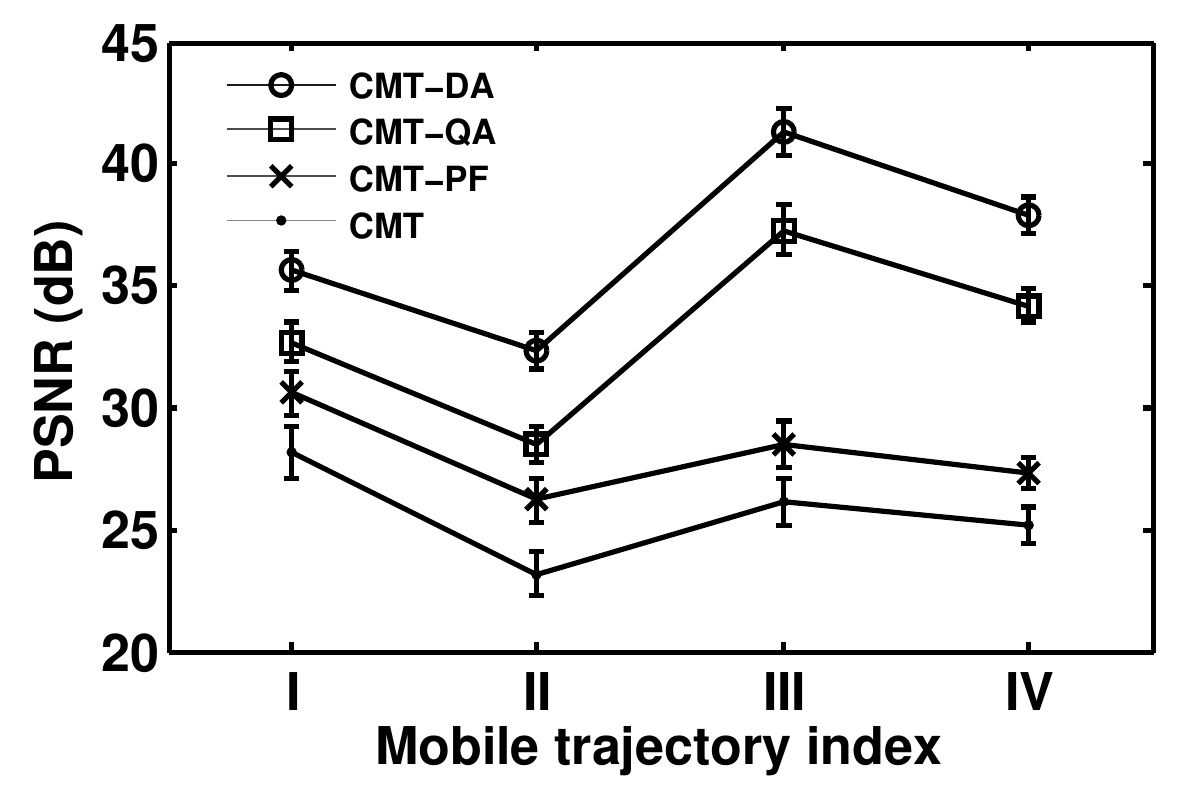}\\
 \centering (a) different trajectories
  \label{4a}
 \end{minipage}%
\begin{minipage}[t]{0.25\linewidth}
\centering
 \includegraphics[width=1\textwidth,keepaspectratio]{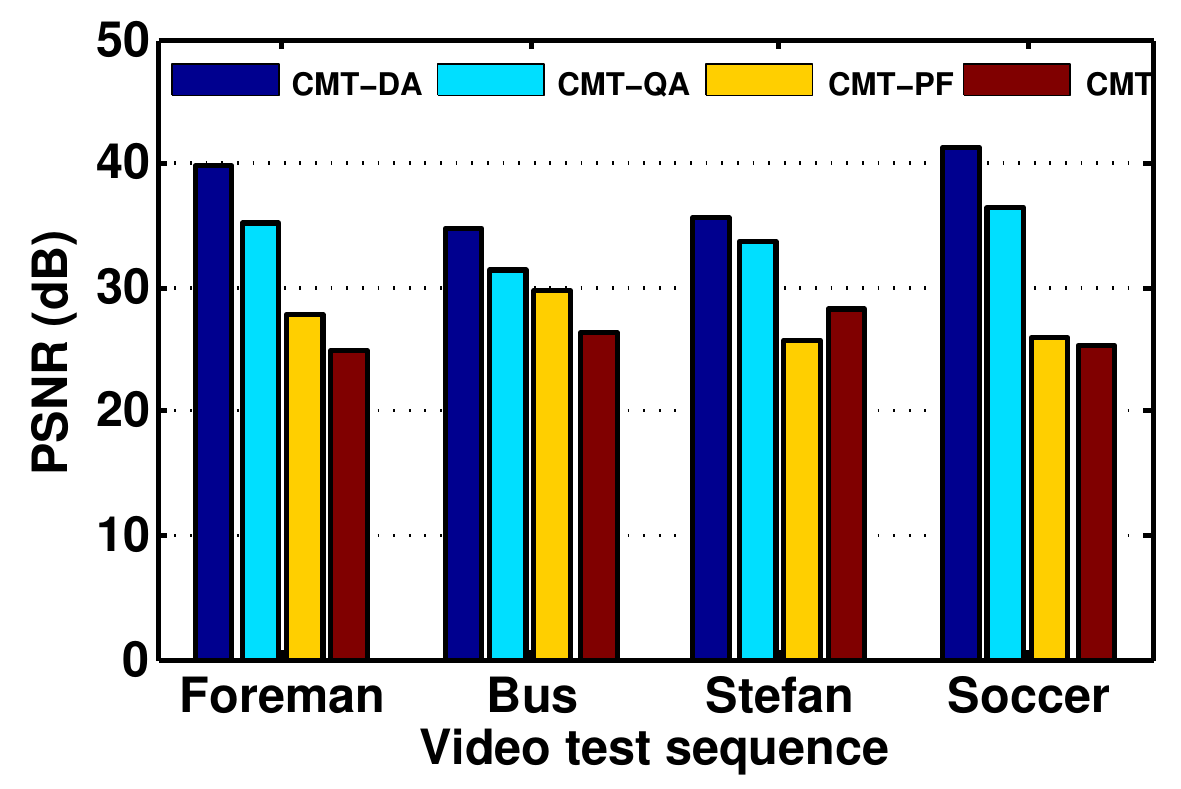}\\
 \centering (b) different sequences
 \label{4b}
\end{minipage}%
\begin{minipage}[t]{0.25\linewidth}
\centering
 \includegraphics[width=1\textwidth,keepaspectratio]{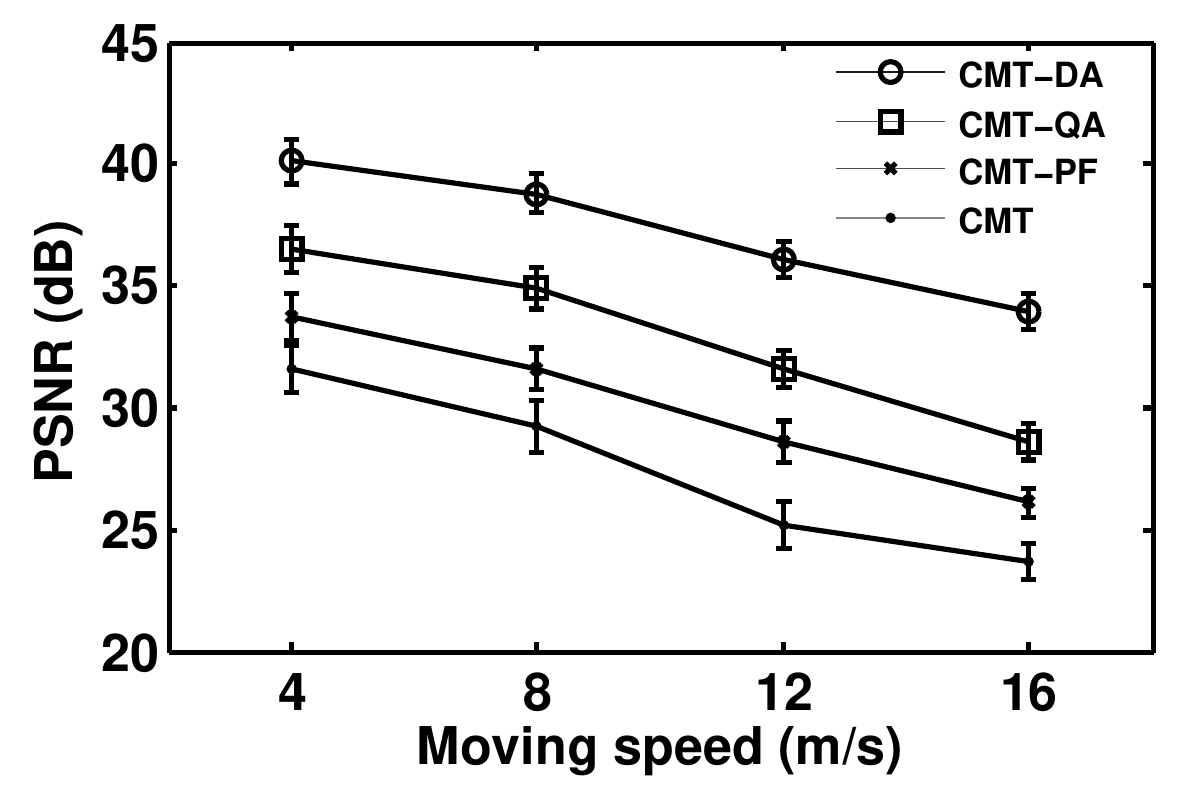}\\
 \centering (c) different speeds
 \label{4b}
\end{minipage}%
\begin{minipage}[t]{0.25\linewidth}
\centering
 \includegraphics[width=1\textwidth,keepaspectratio]{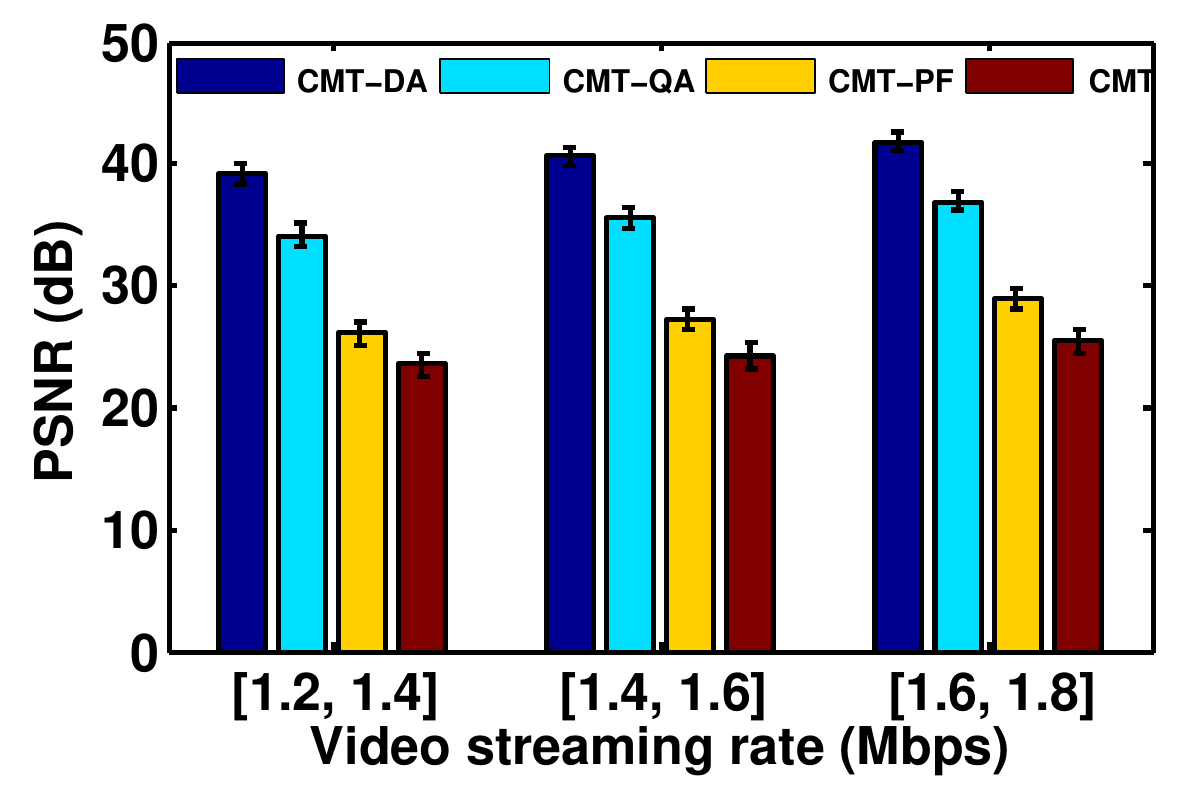}\\
 \centering (d) VBR streaming
 \label{4b}
\end{minipage}%
\caption{Comparison of average PSNR results.}
\end{figure*}
\subsubsection{Reference Schemes}
\begin{itemize}
  \item CMT-QA [5]. The path quality is estimated based on the ratio of chunk delivery time to sending buffer size. CMT-QA distinguishes the network congestions from wireless channel errors with the $RTT-cwnd$ production. The path congestion window size is updated once consecutive losses are detected.
  \item CMT-PF [14]. The path failures are detected via the heartbeats which are sent to the destination with an exponential backoff of RTO after every timeout. If a heartbeat ACK is alive, the path's congestion window is set to be $2$ MTUs.
  \item CMT [13]. The RTX-LOSSRATE is selected as the retransmission policy to minimize the transfer time. Such policy uses the information about loss rate estimated by RTX-CWND.
\end{itemize}
In the emulations, CMT-DA and the reference schemes are implemented in the server and client side to perform online transmission scheduling.
\subsubsection{Performance Metrics}
\begin{itemize}
  \item PSNR [42]. PSNR (Peak Signal-to-Noise Ratio) is the standard metric of objective video quality and is a function of the mean square error between the original and the received video frames. If a video frame experiences transmission or overdue loss, it is considered to be lost and will be concealed by copying from the last received frame before it.
  \item Inter-packet delay. We measure the inter-packet delay of received packets to quantify the jitter of delivered video stream. High jitter values between packets cause bad visual quality (e.g., video glitches and stalls during the display).
  \item Goodput. Goodput is an application-level throughput, i.e., the number of useful information bits successfully received by the destination within the imposed deadline. The amount of data considered excludes protocol overhead bits.
  \item Effective loss rate. As introduced in Section 3.3, the effective loss rate includes both the transmission and overdue losses. 
\end{itemize}

\subsubsection{Emulation Scenarios}
To compare the performance of the reference schemes, we conduct emulations in the mobile scenario with trajectories indexed from I to IV as shown in Fig. 3. The four mobile trajectories represent the different access options for the mobile user in the integrated heterogeneous wireless networks. The mobile client requests to the server through a wireless interface and constructs the connection whenever it moves into the coverage. The initial moving speed of the client is set to be $2$ m/s. 

For the confidence results, we repeat each set of emulations with different video sequences more than $20$ times and obtain the average results with a $95\%$ confidence interval. The microscopic and mobility results are presented with the measurements of finer granularity.

\begin{figure*}[htbp]
        \centering
        \begin{minipage}[t]{0.25\linewidth}
         \centering
         \includegraphics[width=1\textwidth,keepaspectratio]{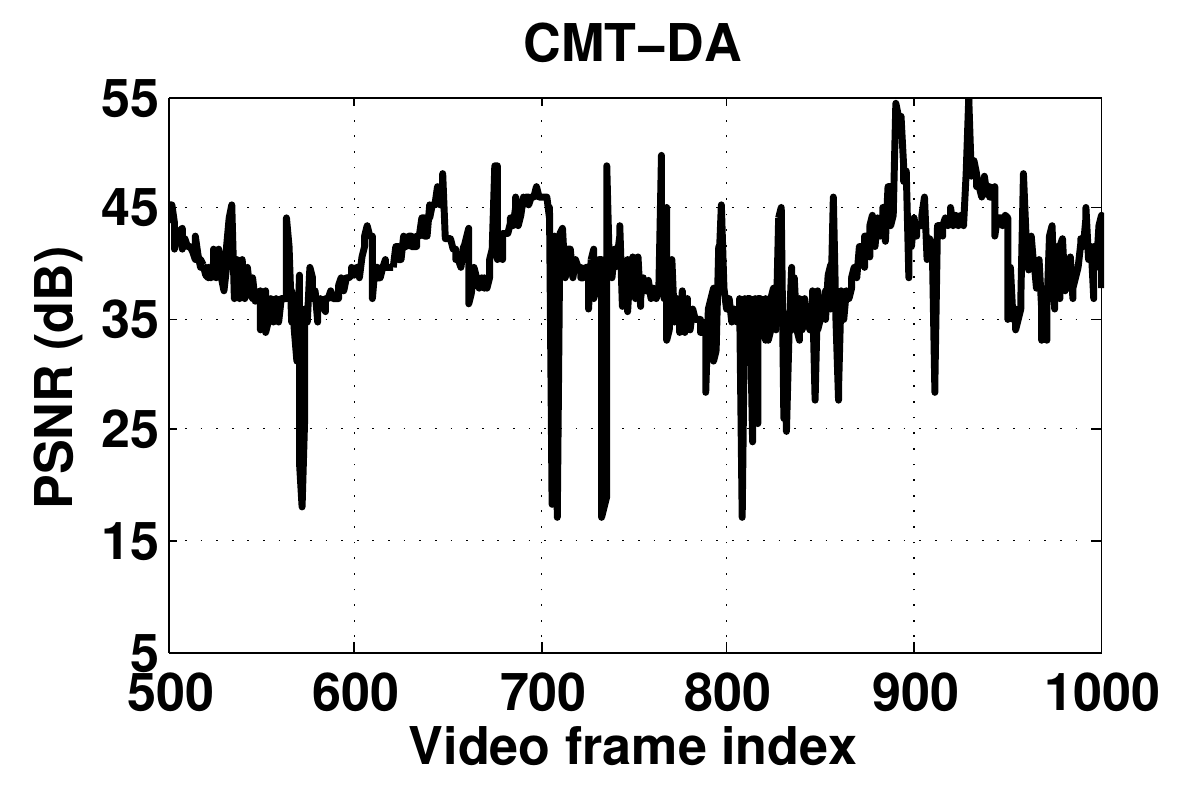}
            \centering
         \label{fig:side:a}
        \end{minipage}%
        \begin{minipage}[t]{0.25\linewidth}
         \centering
        \includegraphics[width=1\textwidth,keepaspectratio]{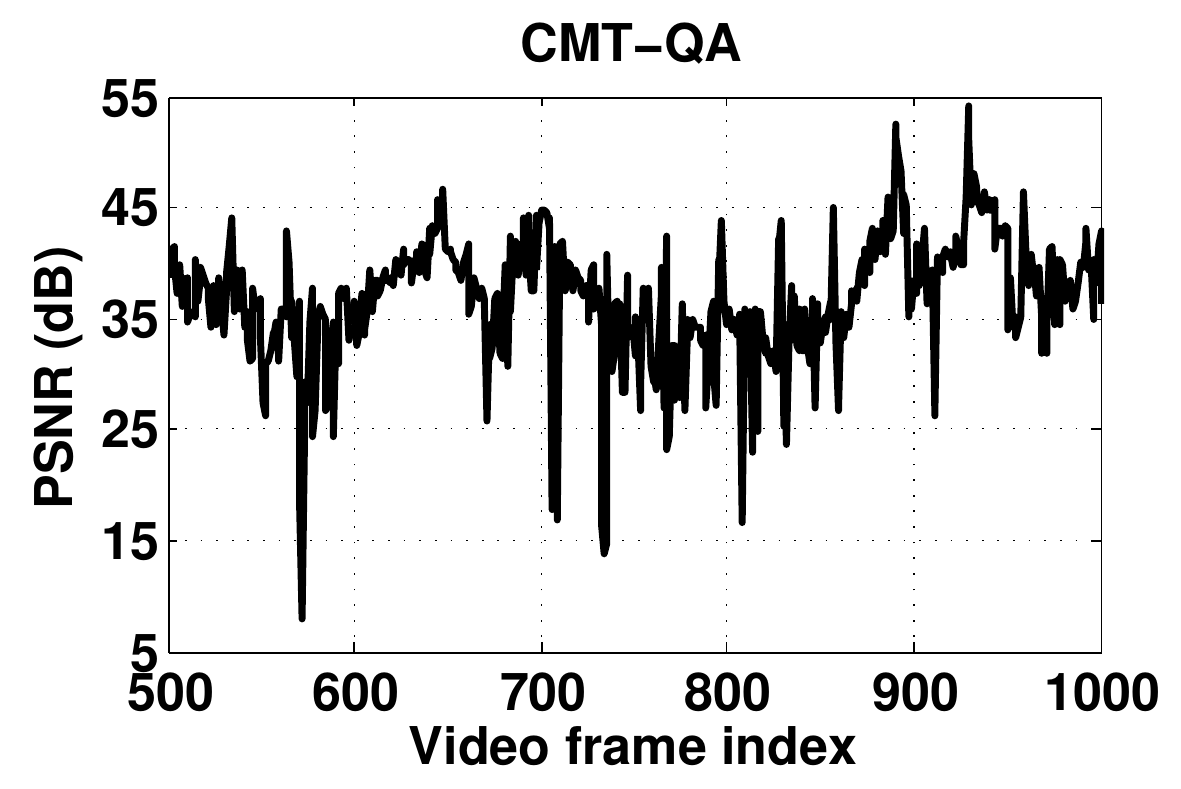}
         \label{fig:side:b}
        \end{minipage}%
        \begin{minipage}[t]{0.25\linewidth}
         \centering
        \includegraphics[width=1\textwidth,keepaspectratio]{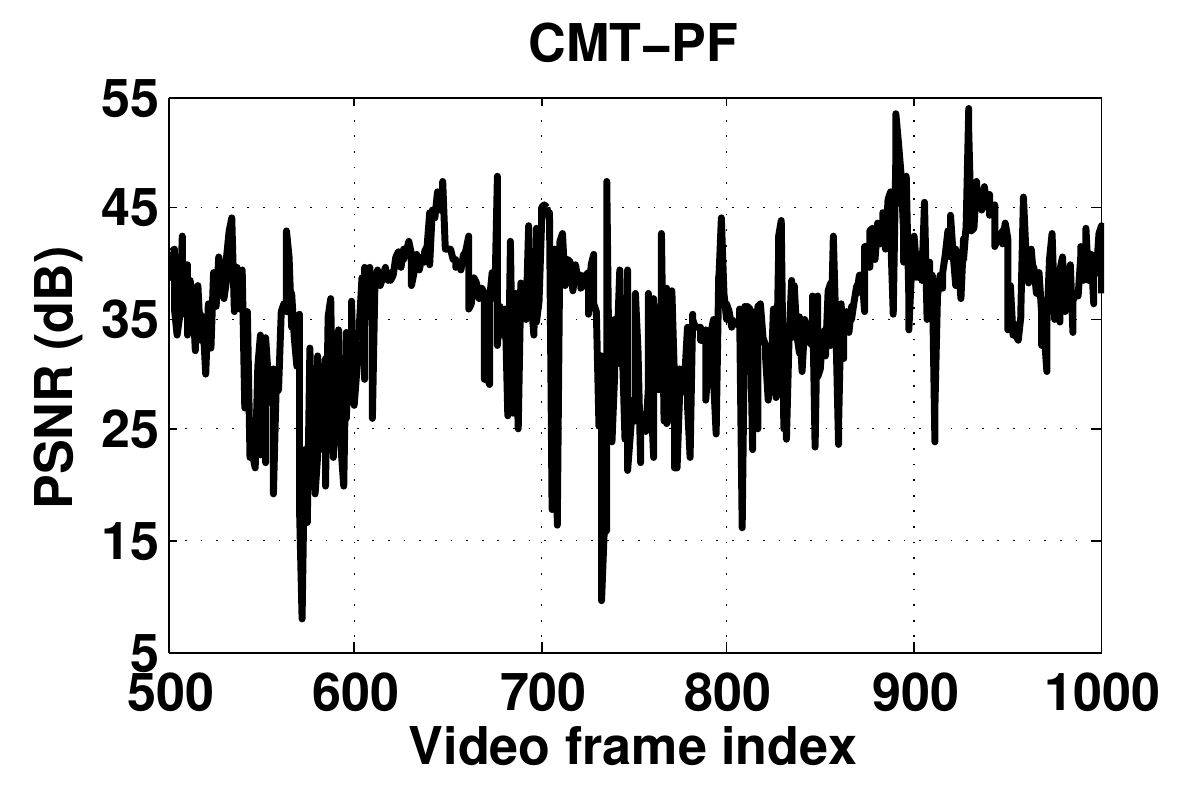}
         \label{fig:side:c}
        \end{minipage}%
         \begin{minipage}[t]{0.25\linewidth}
         \centering
        \includegraphics[width=1\textwidth,keepaspectratio]{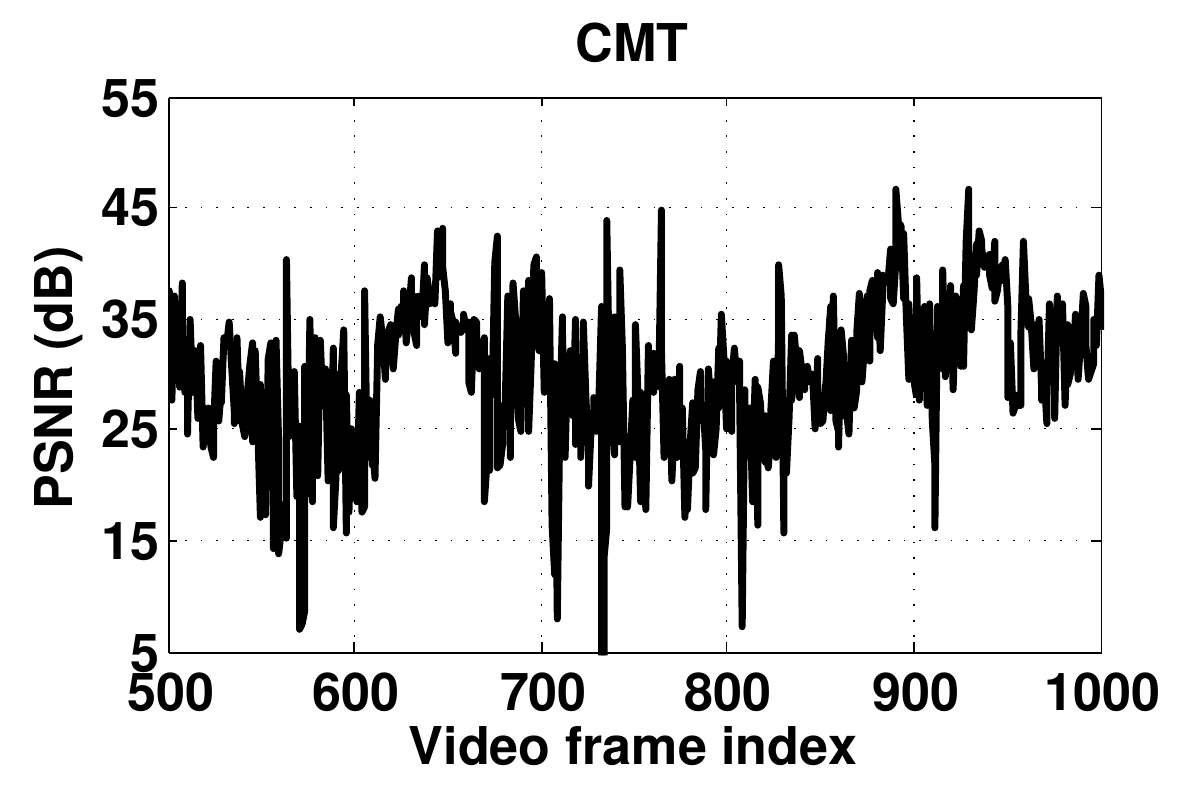}
         \label{fig:side:c}
        \end{minipage}%
         \caption{Comparison of PSNR per video frame measured from the \emph{Soccer} sequence.}
        \label{7}
        \end{figure*}
        \begin{figure*}[htbp]
        \centering
        \begin{minipage}[t]{0.24\linewidth}
         \centering
         \includegraphics[width=1\textwidth,keepaspectratio]{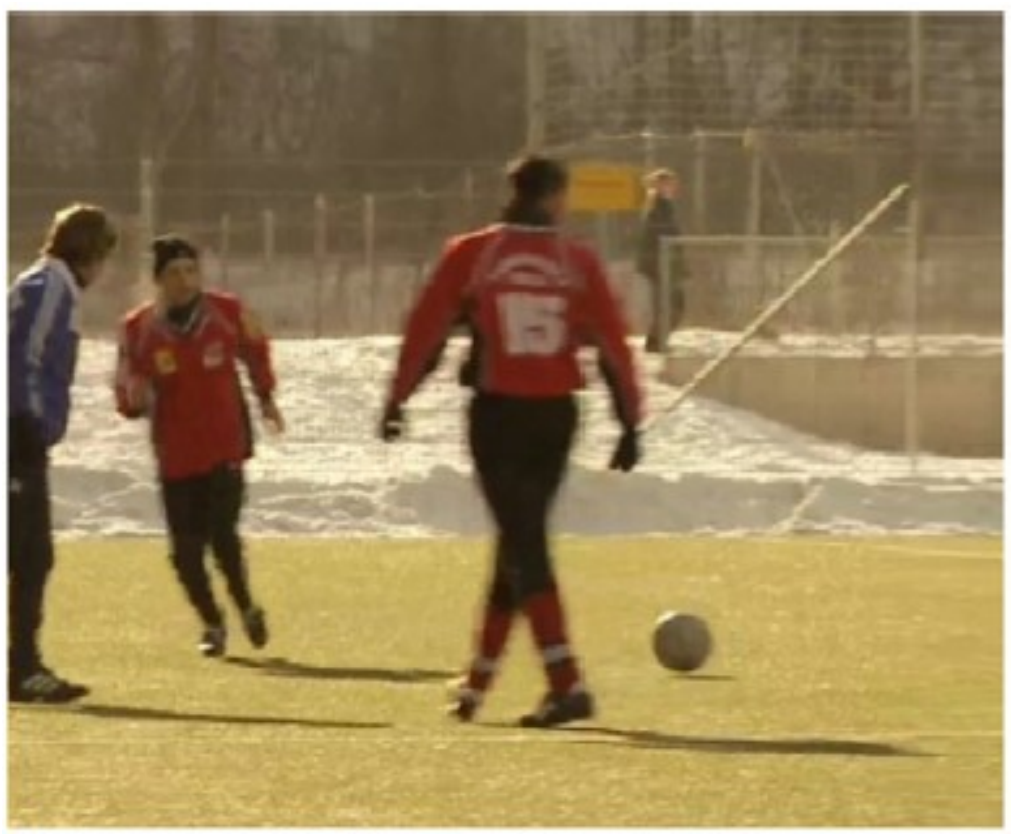}\\
         (a) CMT-DA
            \centering
         \label{fig:side:a}
        \end{minipage}\,
        \begin{minipage}[t]{0.24\linewidth}
         \centering
        \includegraphics[width=1\textwidth,keepaspectratio]{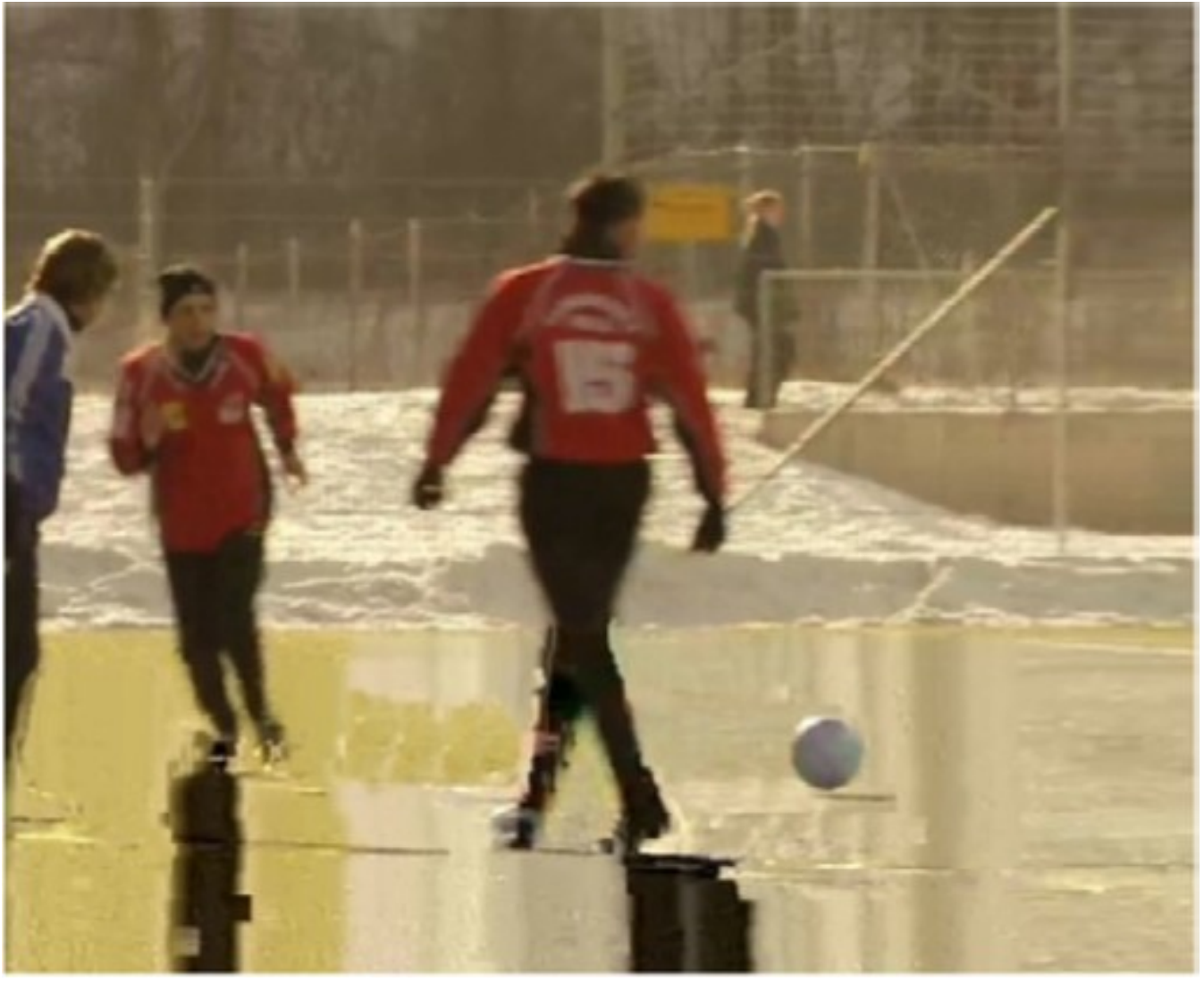}\\
         (b) CMT-QA
         \label{fig:side:b}
        \end{minipage}\,
        \begin{minipage}[t]{0.24\linewidth}
         \centering
        \includegraphics[width=1\textwidth,keepaspectratio]{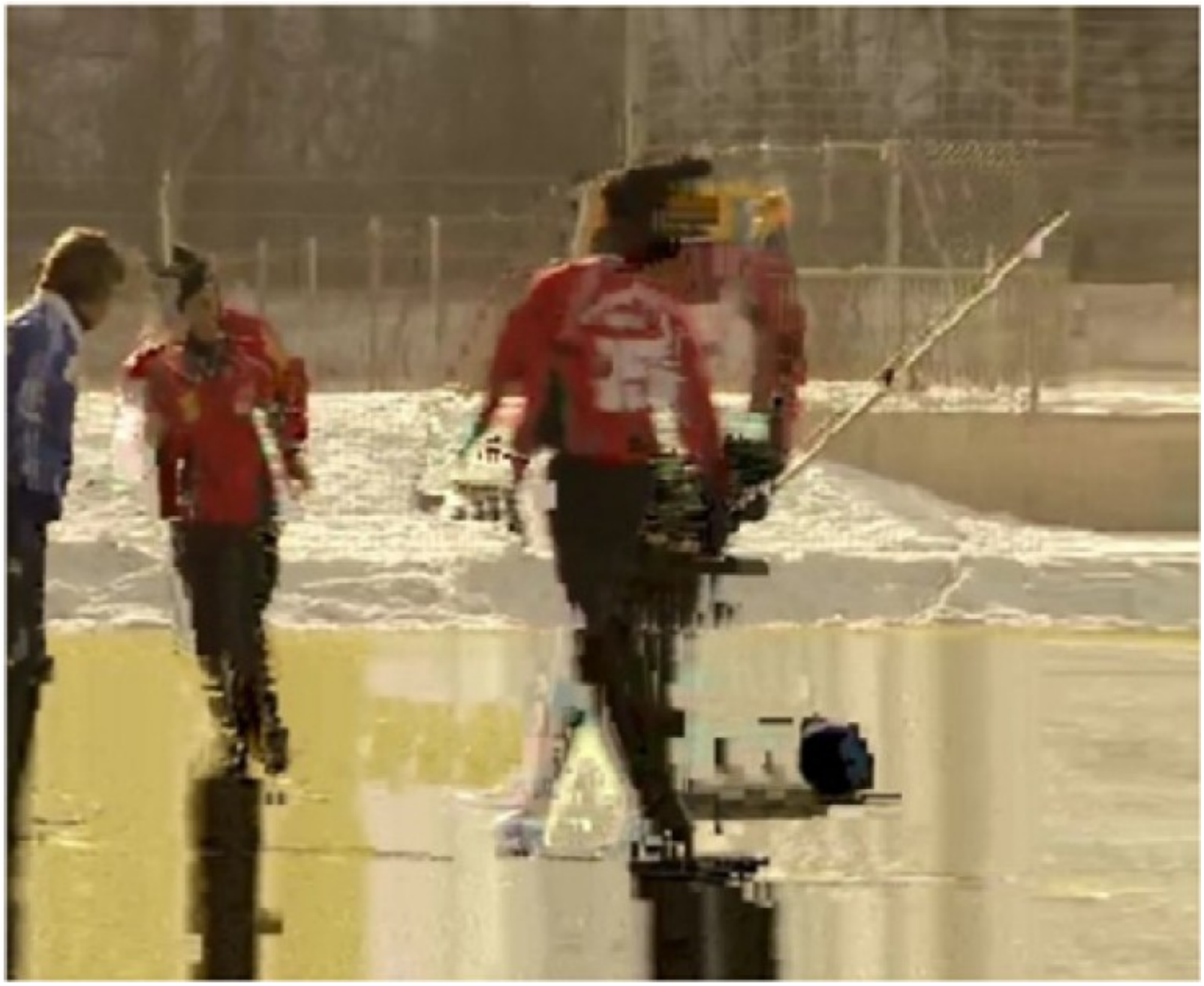}\\
         (c) CMT-PF
         \label{fig:side:c}
        \end{minipage}\,
         \begin{minipage}[t]{0.24\linewidth}
         \centering
        \includegraphics[width=1\textwidth,keepaspectratio]{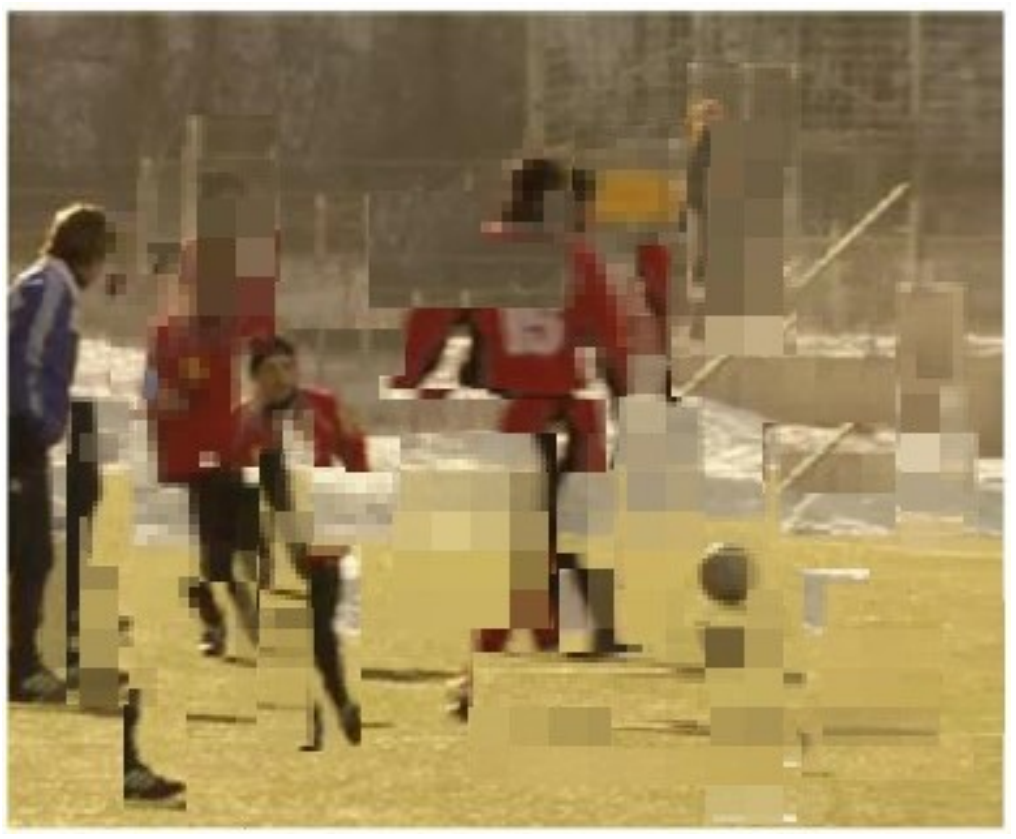}\\
         (d) CMT
         \label{fig:side:c}
        \end{minipage}%
         \caption{Comparison of subjective video quality measured from the \emph{Soccer} sequence.}
        \label{7}
        \end{figure*}
\subsection{Evaluation Results}
We firstly depict the channel status information obtained by the implemented path monitoring algorithm in Fig. 4. It can be observed that HSDPA supports relatively stable link while the available bandwidth of WiMAX and WLAN experiences frequent fluctuations. It is well-known that cellular networks exhibit better performance in sustaining user mobility than WiMAX and WLAN but provide a lower peak data rate [6][37].

\subsubsection{PSNR}
Fig. 5a plots the average PSNR values and confidence intervals of all the competing CMT schemes. We can see that CMT-DA achieves higher PSNR values with lower variations than the reference schemes. The proposed CMT-DA takes into account the delay constraint imposed by the video applications while the reference schemes are unaware of the video traffic information. Fig. 5b show the average PSNR values measured from each video test sequence while the mobile client is moving along Trajectory III. We also measure the PSNR results under different moving speeds along the Trajectory III. As shown in Fig. 5c, CMT-DA outperforms the reference schemes in sustaining high speed. The performance comparison for Variable Bit-Rate (VBR) video streaming is presented in Fig. 5d. Expectedly, CMT-DA achieves better performance for taking into account the bit rate variability.

In order to have a microscopic view of the results, we plot the PSNR per video frame indexed from $500$ to $1000$ measured from the \emph{Soccer} sequence in Fig. 6. The variations are probably caused by the channel fading or injected background traffic. Similar subjective results are observed when we watch streaming video on the receiver side. Indeed, we notice more frequent glitches and stalls with the reference schemes, while much smoother streaming is obtained with CMT-DA. To display the perceived quality, a comparison of the $623$th received video frame is illustrated in Fig. 7. It can be observed that CMT-DA achieves better perceived video quality while the received video frames of CMT-PF and CMT are seriously damaged.

\subsubsection{Inter-packet Delay}
\begin{figure}[htbp]
\centering
\begin{minipage}[t]{0.5\linewidth}
\centering
 \includegraphics[width=1\textwidth,keepaspectratio]{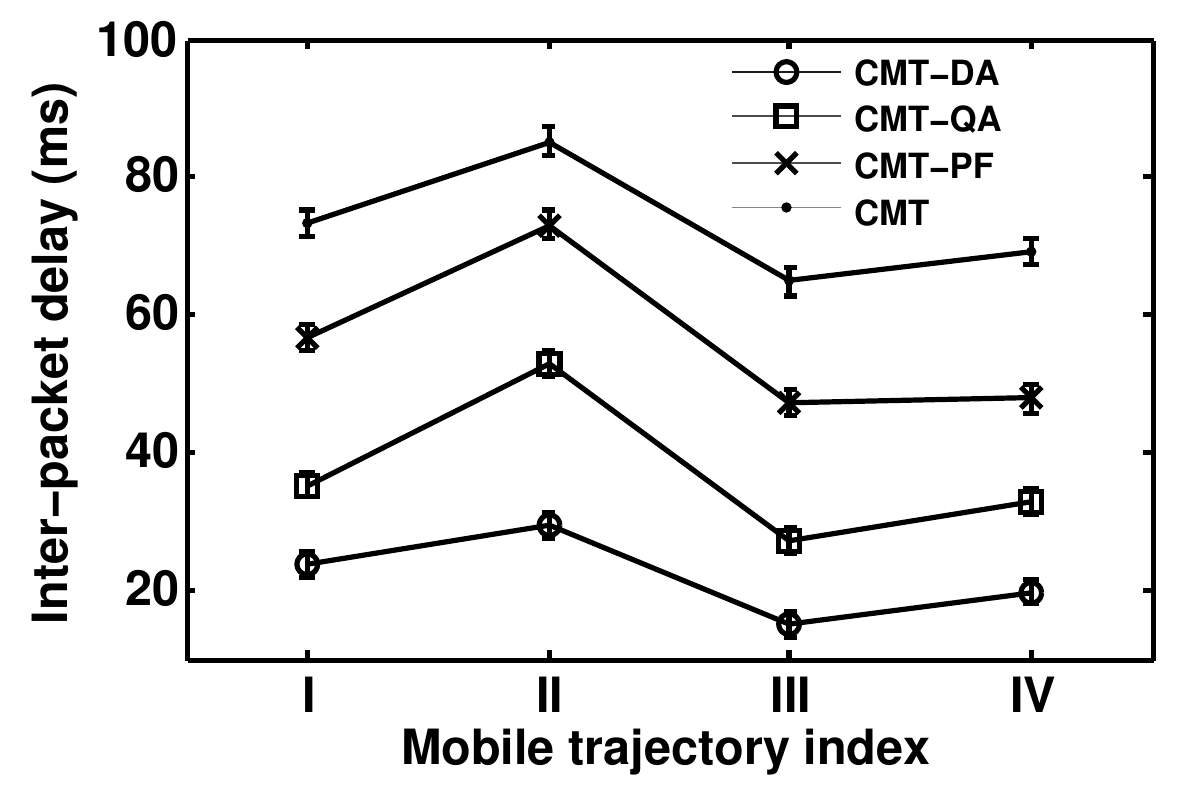}\\
 \centering (a) average delay
  \label{4a}
 \end{minipage}%
\begin{minipage}[t]{0.5\linewidth}
\centering
 \includegraphics[width=1\textwidth,keepaspectratio]{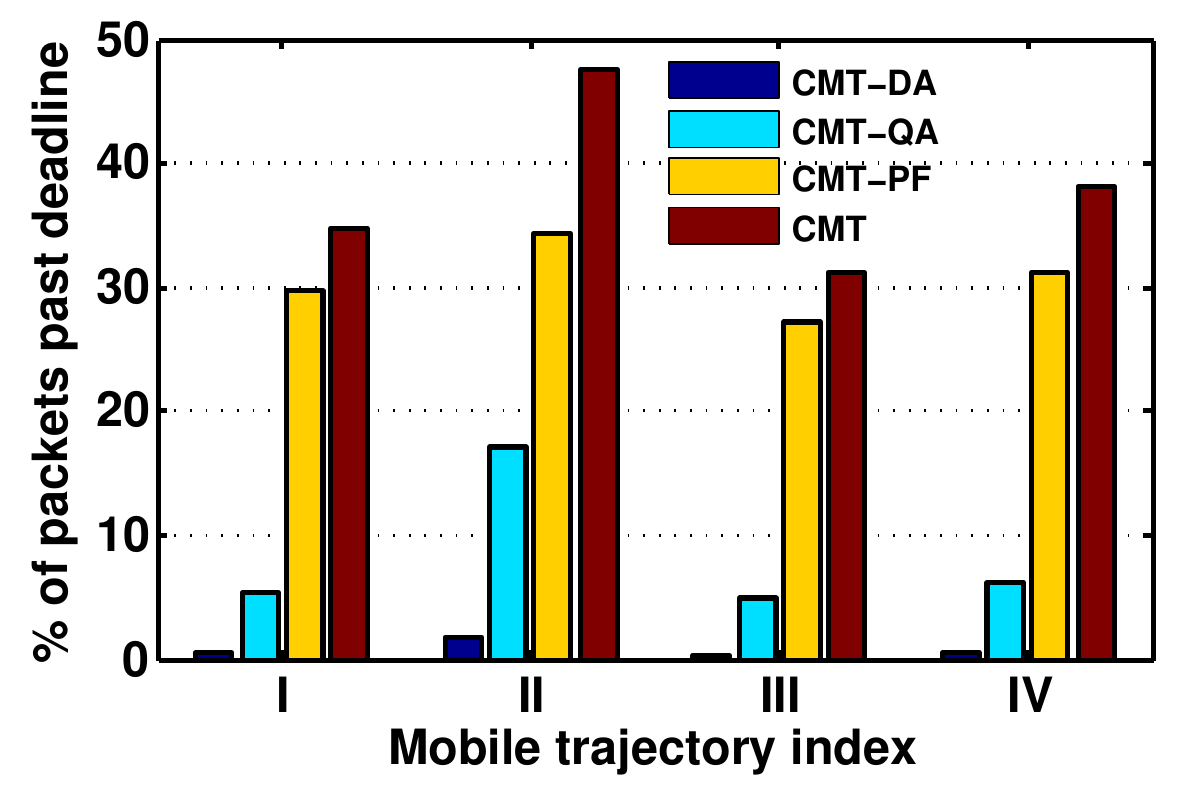}\\
 \centering (b) $\%$ of overdue packets
 \label{4b}
\end{minipage}%
\caption{Comparison of inter-packet delay and ratio of overdue packets.}
\end{figure}
The average inter-packet delay of the four competing schemes with respect to different mobile trajectories is shown in Fig. 8a. The results' pattern is almost opposite to that presented in Fig. 5a. As a rule of thumb, larger end-to-end delays result in lower video quality in real-time applications. CMT-DA achieves significantly lower delays than the reference schemes and Fig. 8b shows the resulting ratio of overdue packets in all the simulation scenarios. To have a close-up view of the results, we plot the Cumulative Distribution Functions (CDF) of inter-packet delay with respect to different trajectories in Fig. 9. As we can observe from Fig. 9a, c and d that, CMT-DA guarantees more than $60\%$ of the packets having inter-arrival time less than $20$ ms. This indicates significantly fewer interruptions and stalls perceived by the end users.
\begin{figure*}[htbp]
        \centering
        \begin{minipage}[t]{0.25\linewidth}
         \centering
         \includegraphics[width=1\textwidth,keepaspectratio]{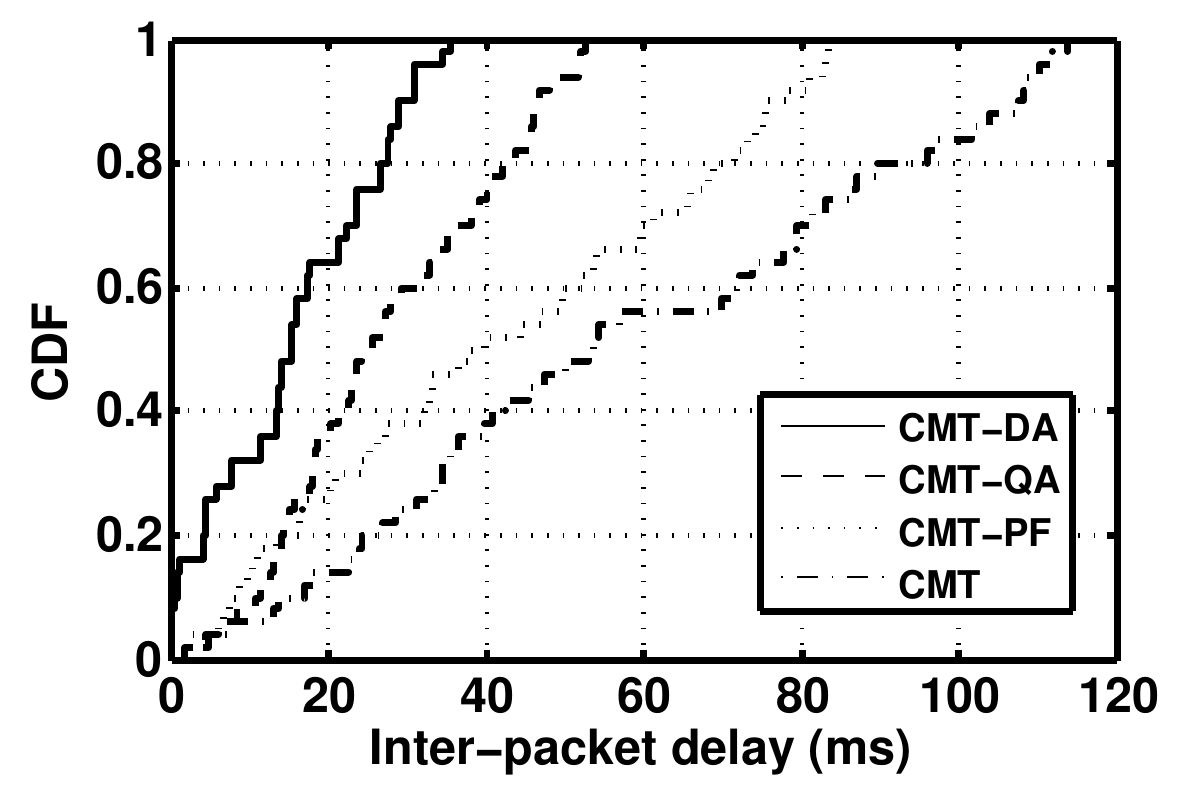}\\
         (a) Trajectory I
            \centering
         \label{fig:side:a}
        \end{minipage}%
        \begin{minipage}[t]{0.25\linewidth}
         \centering
        \includegraphics[width=1\textwidth,keepaspectratio]{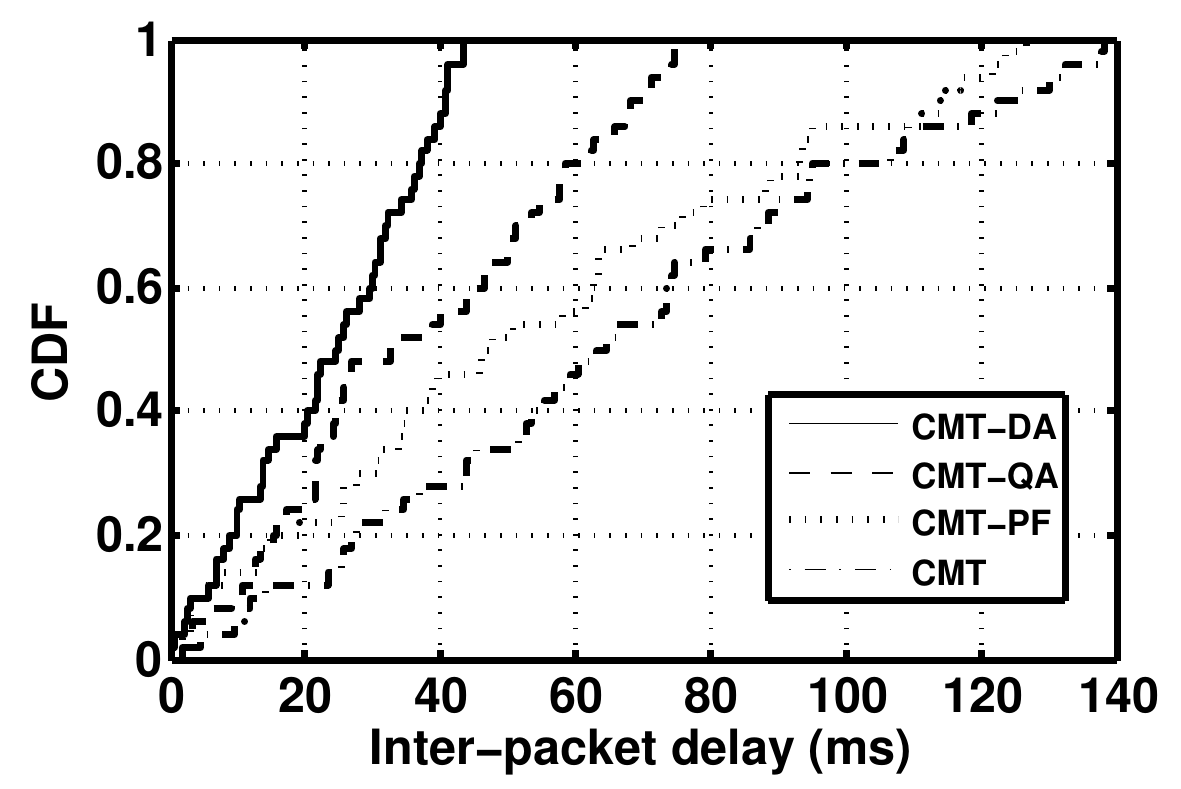}\\
         (b) Trajectory II
         \label{fig:side:b}
        \end{minipage}%
        \begin{minipage}[t]{0.25\linewidth}
         \centering
        \includegraphics[width=1\textwidth,keepaspectratio]{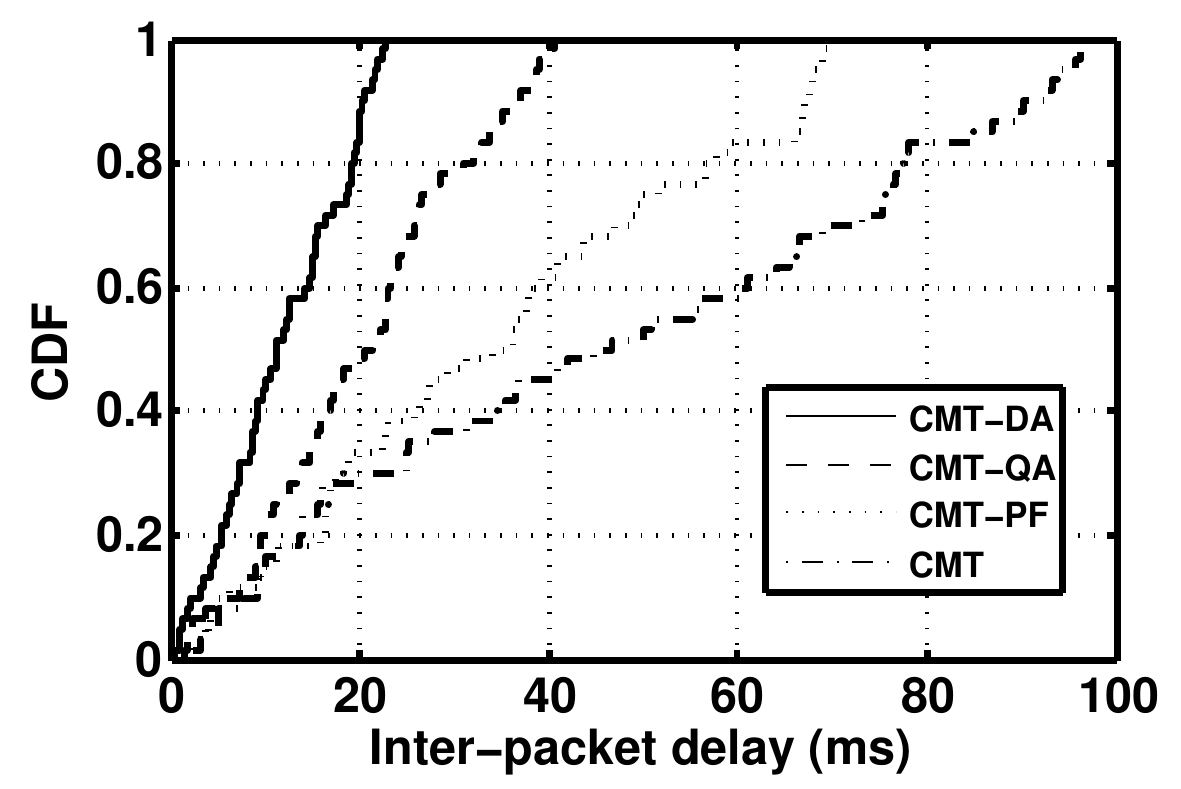}\\
         (c) Trajectory III
         \label{fig:side:c}
        \end{minipage}%
         \begin{minipage}[t]{0.25\linewidth}
         \centering
        \includegraphics[width=1\textwidth,keepaspectratio]{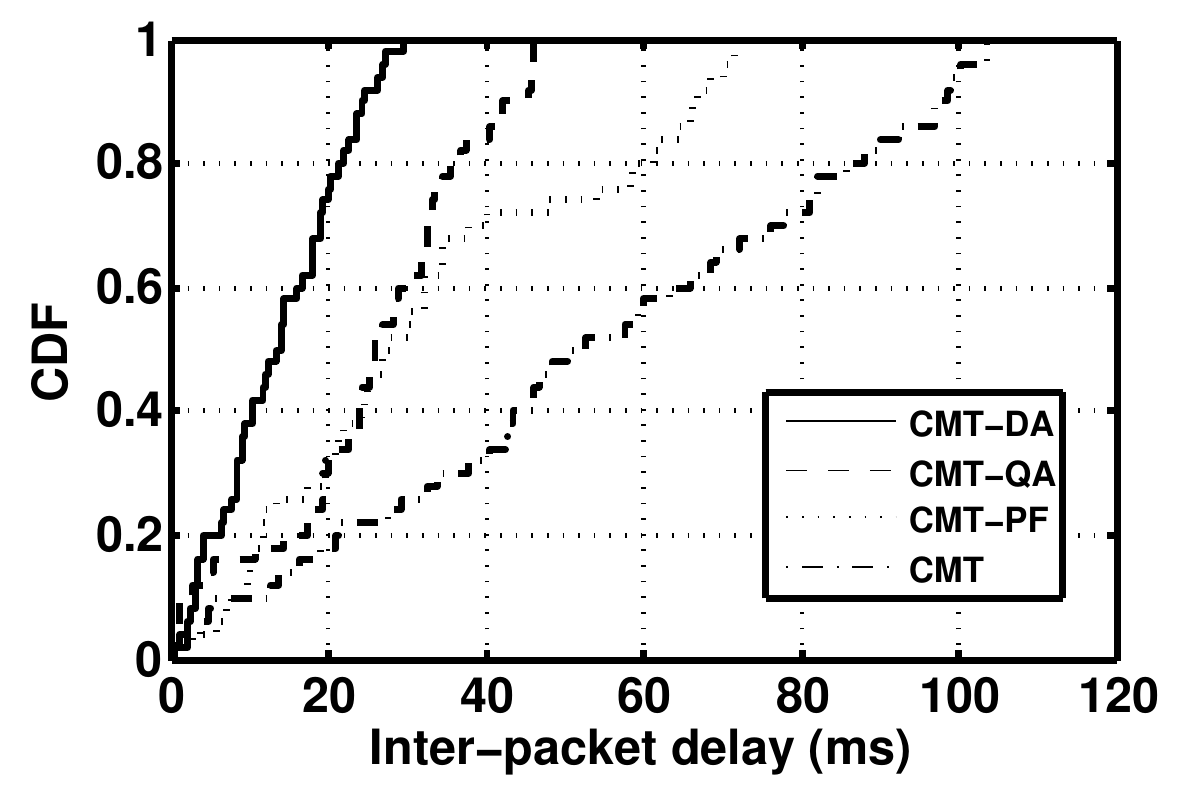}\\
         (d) Trajectory IV
         \label{fig:side:c}
        \end{minipage}%
         \caption{CDF of inter-packet arrival time in all the emulations.}
        \label{7}
\end{figure*}

\begin{figure}[htbp]
\centering
 \includegraphics[width=0.4\textwidth,keepaspectratio]{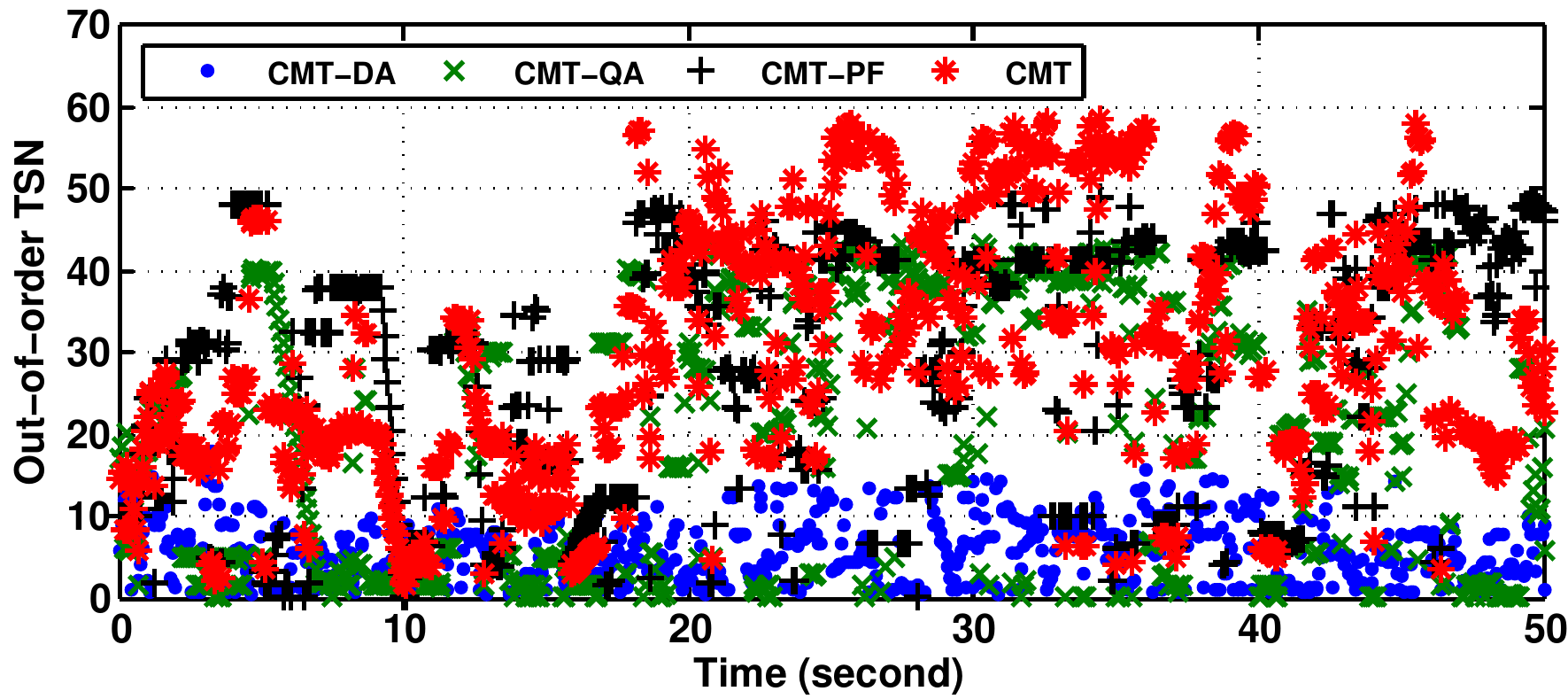}%
 \caption{Comparison of out-of-order Transmission Sequence Number (TSN).}\label{fig1}
 \end{figure}

Although the inter-packet delay is an important metric for real-time applications, the probability of out-of-order chunks is also a critical problem in the context of CMT because SCTP has the responsibility to reorder the chunks for restoring the input traffic. Fig. 10 depicts the out-of-order chunks for the evaluated schemes. The out-of-order TSN metric used in this experiment is measured by the offset between the TSNs of two consecutively received data chunks (the difference between the TSN of the current data chunk and that of the latest received data chunk). The results reflect the competing schemes' capability to overcome the path asymmetry. Both the CMT-DA and CMT-QA periodically estimate the path status and report to the sender side. The latest information can increase the accuracy of predicting the end-to-end delay. As is shown in the figure that CMT-PF and CMT induce more reordering chunks than CMT-DA and CMT-QA. Note that the out-of-order problem not only induces the additional recovery time at the destination but also results in receiving buffer blocking. The problem becomes more critical when the receiving buffer size is limited in mobile devices.

\subsubsection{Goodput}
\begin{figure}[htbp]
\centering
\begin{minipage}[t]{0.5\linewidth}
\centering
 \includegraphics[width=1\textwidth,keepaspectratio]{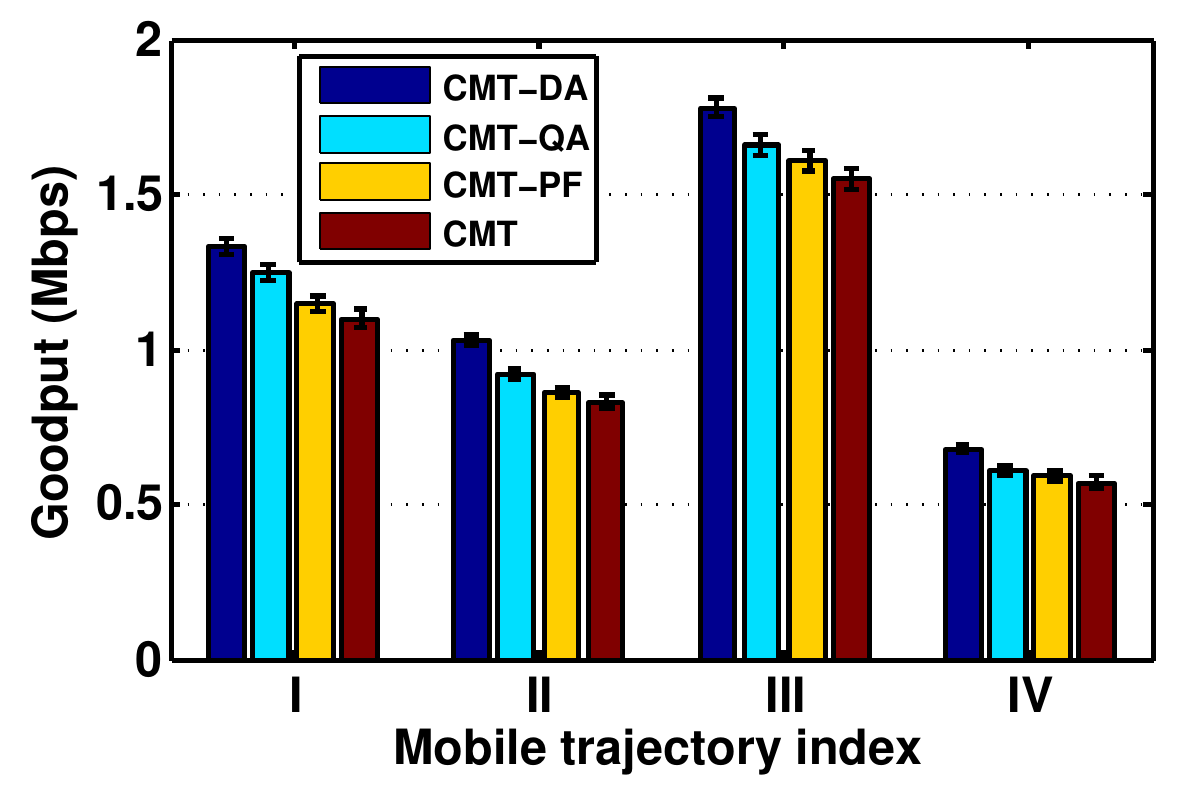}\\
 \centering (a) average goodput
  \label{4a}
 \end{minipage}%
\begin{minipage}[t]{0.5\linewidth}
\centering
 \includegraphics[width=1\textwidth,keepaspectratio]{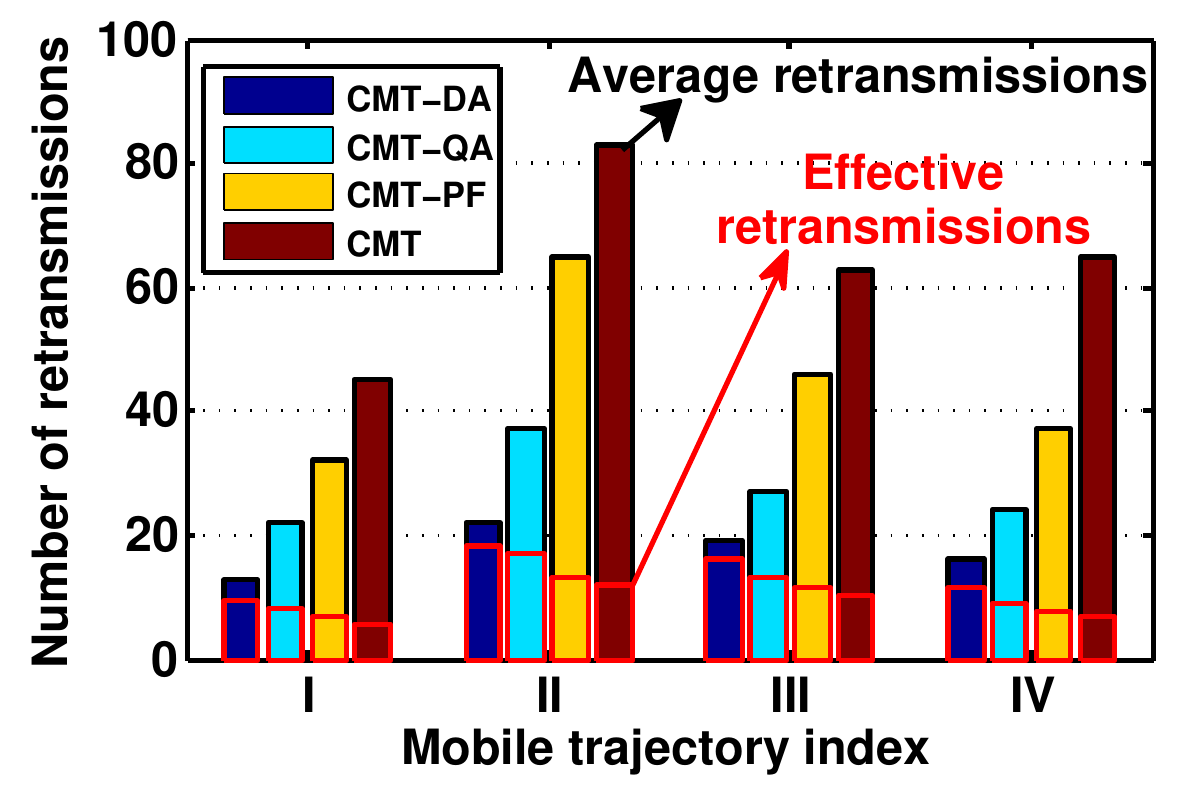}\\
 \centering (b) retransmissions
 \label{4b}
\end{minipage}%
\caption{Comparison of average goodput and retransmission results.}
\end{figure}

Fig. 11a plots the average goodput for the reference schemes in the emulations along different trajectories. The gap between CMT-DA and other solutions become larger in mobile trajectory III. It is similar to the results' pattern shown in Fig. 5a. It is well-know that goodput differs from throughput with regard to the imposed deadline. \emph{The video PSNR or subjective quality is not only correlated with the goodput, but also depends on the weight of the lost video frames, statistics of scene contents, etc.} The unnecessary retransmissions may increase the effective loss rate and degrade the goodput performance of input video streaming. We plot the number of average and effective retransmissions in Fig. 11b. It can be observed that the proposed delay and loss controlled retransmission policy outperforms the conventional CWND/LOSSRATE schemes. The effective retransmission is improved by only re-sending the data chunks which are estimated to arrive at the destination.

\begin{figure*}[htbp]
        \centering
        \begin{minipage}[t]{0.25\linewidth}
         \centering
         \includegraphics[width=1\textwidth,keepaspectratio]{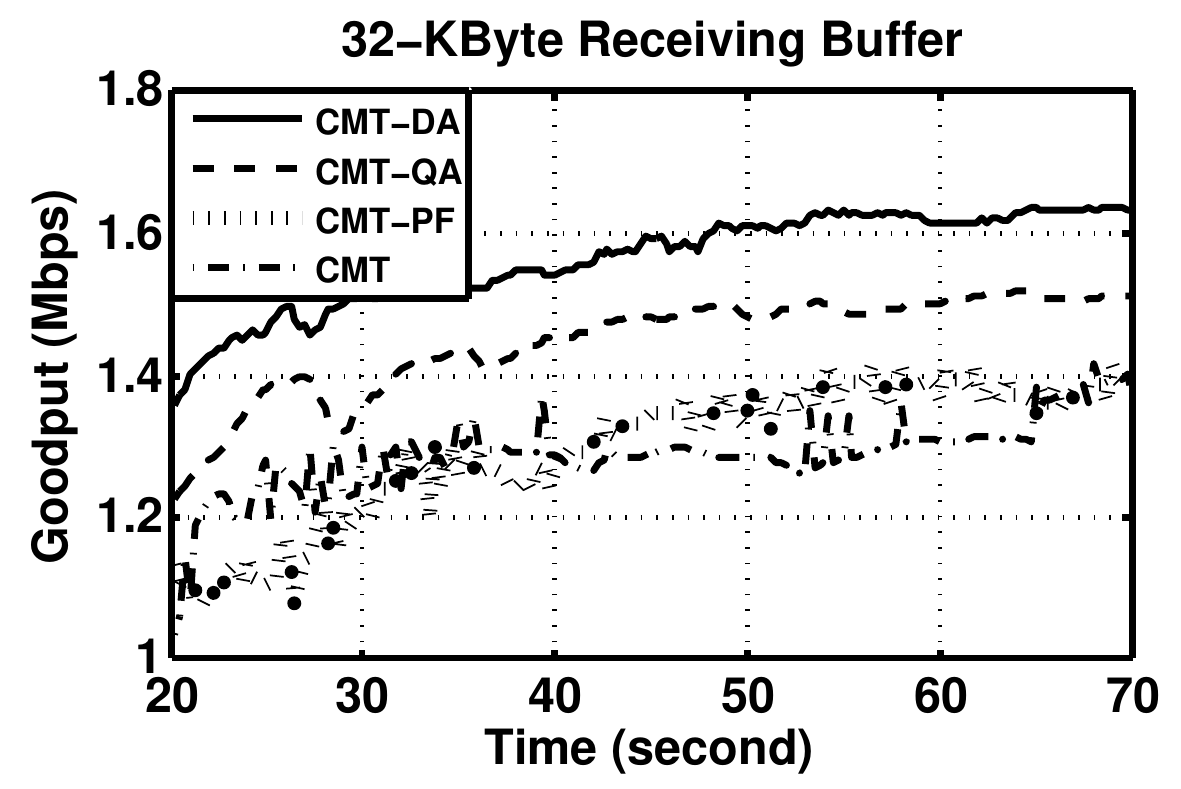}
            \centering
         \label{fig:side:a}
        \end{minipage}%
        \begin{minipage}[t]{0.25\linewidth}
         \centering
        \includegraphics[width=1\textwidth,keepaspectratio]{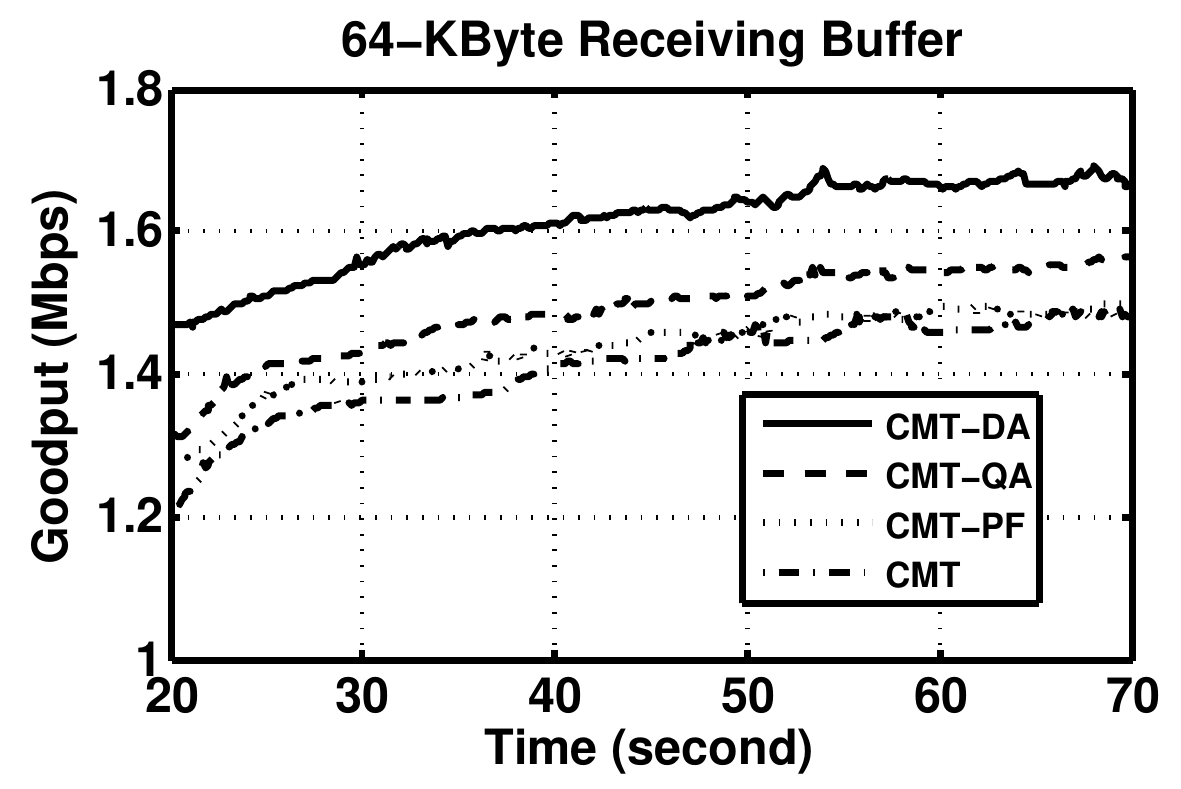}
         \label{fig:side:b}
        \end{minipage}%
        \begin{minipage}[t]{0.25\linewidth}
         \centering
        \includegraphics[width=1\textwidth,keepaspectratio]{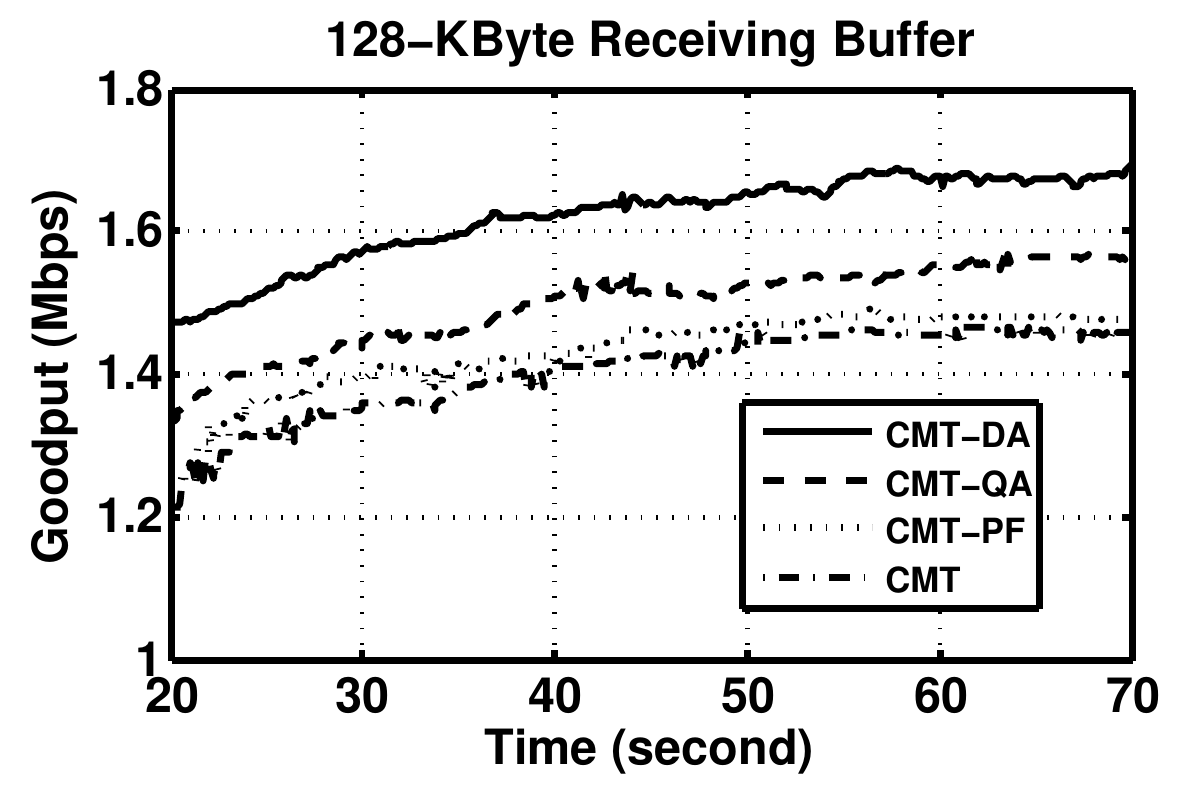}
         \label{fig:side:c}
        \end{minipage}%
         \begin{minipage}[t]{0.25\linewidth}
         \centering
        \includegraphics[width=1\textwidth,keepaspectratio]{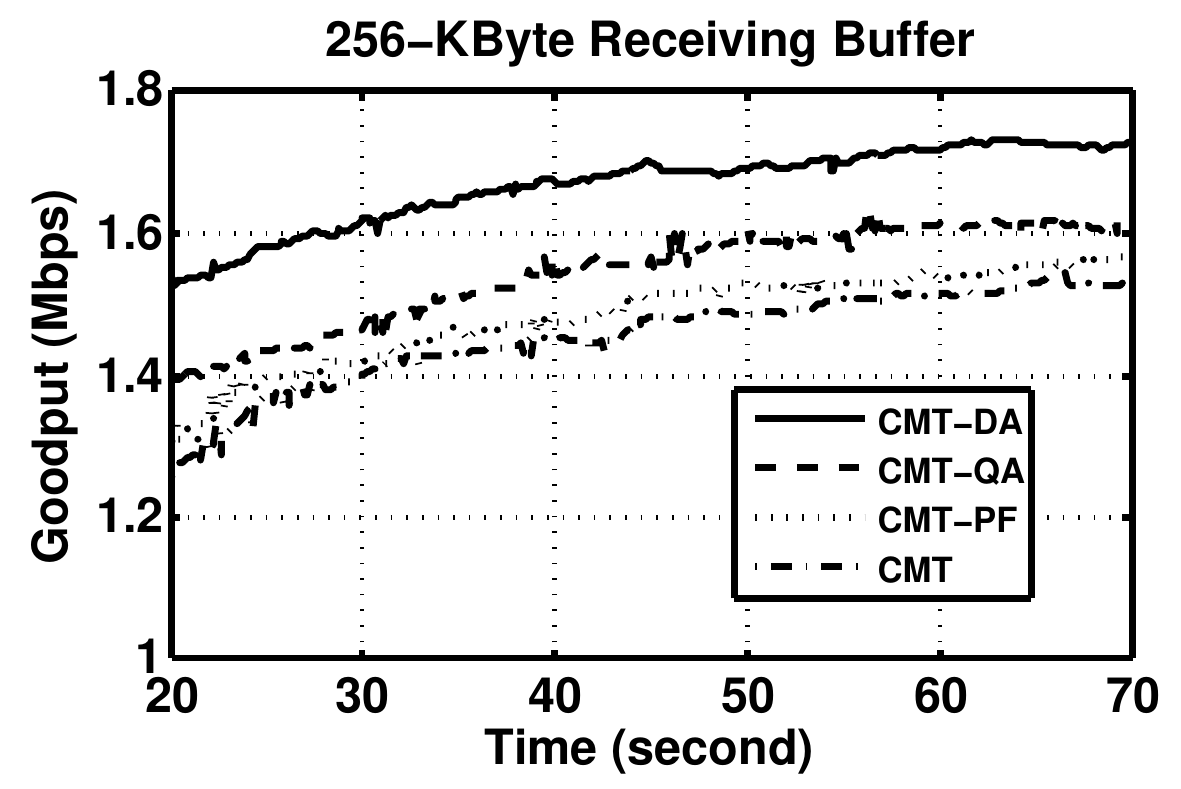}
         \label{fig:side:c}
        \end{minipage}%
         \caption{Comparison of instantaneous goodput values during the interval of $[20,70]$ seconds.}
        \label{7}
\end{figure*}
Fig. 12 compares the instantaneous goodput when delivering streaming video with receiving buffer size of $32$, $64$, $128$, and $256$ KBytes in the emulation of Trajectory III. Due to the path probing and slow start process, the observed goodput values of all the schemes are close to each other at the beginning. Therefore, we compare the instantaneous values measured during the interval of $[20,70]$ seconds. As the raw data is often very noisy, we present the results with finer granularity using the moving average. Compare with the reference schemes, CMT-DA is able to better exploit the available channel resources and increase the goodput more rapidly to the peak value. Generally, a larger receiving buffer is able to store more receiving packets and tolerate longer inter-packet delays. 

\subsubsection{Effective Loss Rate}
\begin{figure}[htbp]
\centering
\begin{minipage}[t]{0.5\linewidth}
\centering
 \includegraphics[width=1\textwidth,keepaspectratio]{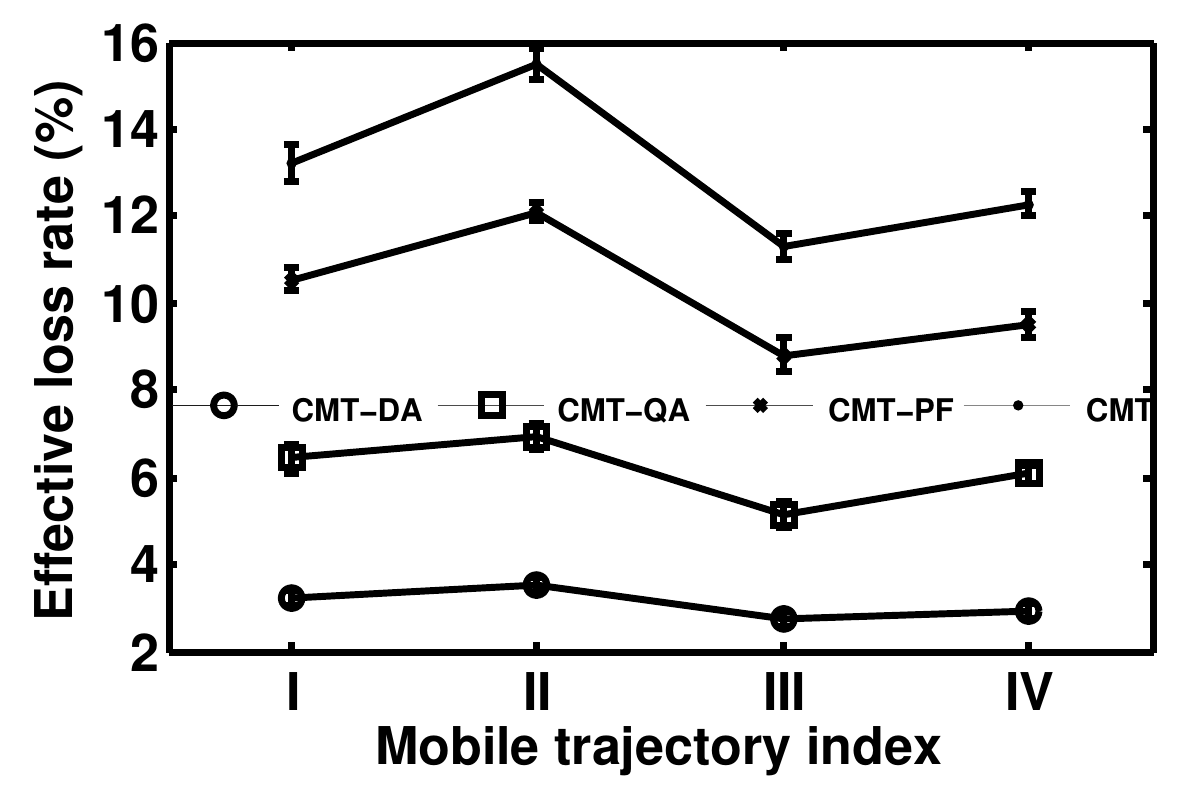}\\
 \centering (a) average values
  \label{4a}
 \end{minipage}%
\begin{minipage}[t]{0.5\linewidth}
\centering
 \includegraphics[width=1\textwidth,keepaspectratio]{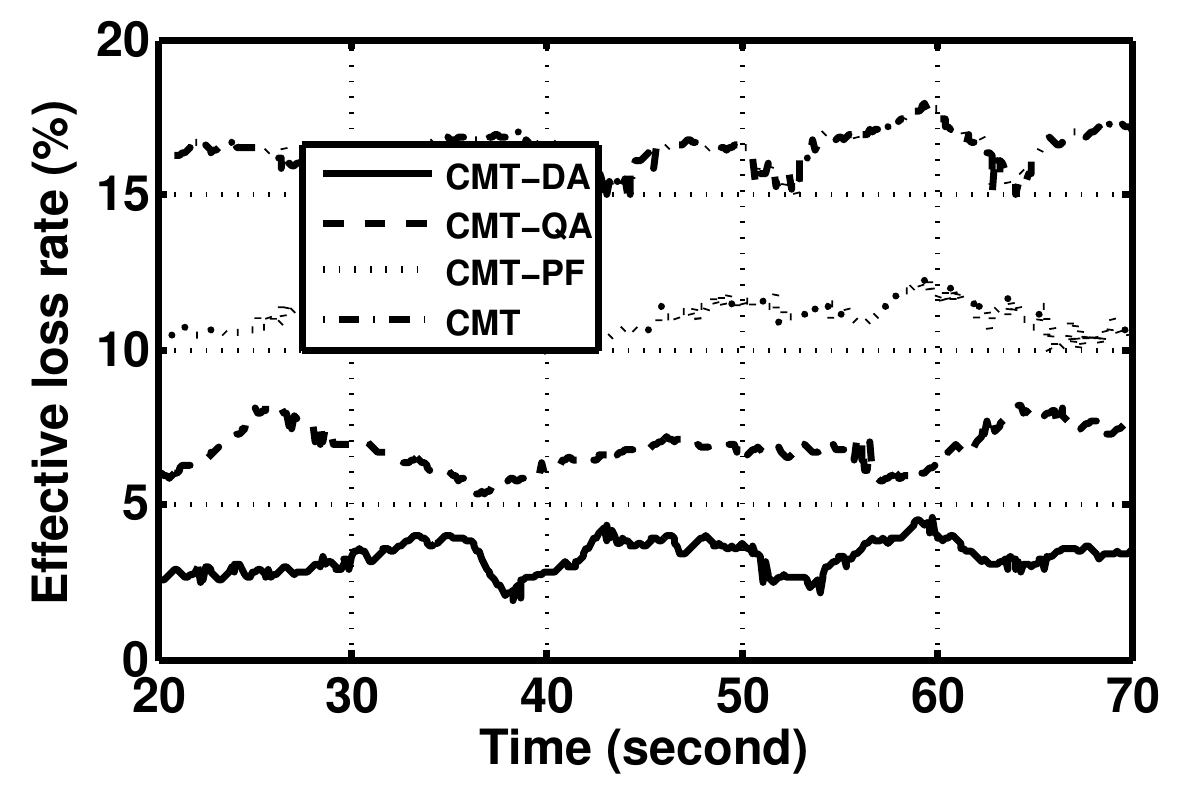}\\
 \centering (b) instantaneous values
 \label{4b}
\end{minipage}%
\caption{Comparison of average effective loss rate and the instantaneous values.}
\end{figure}
Fig. 13a plots the average effective loss rates of all the competing schemes in different mobile scenarios. The effective loss rate is related to the video packet receptions. The more packets the multihomed terminal receives within the deadline, the lower effective loss rate is observed. As is expected, CMT-DA outperforms the competing schemes due to its superiority in both delay and loss performance. CMT-PF outperforms the CMT because it is able to pick out the paths with lower quality and mark it as a `path failure'. Fig. 13b sketches the evolutions of the effective loss rate during the interval of $[20,70]$ seconds. The presented results are refined from the sampled values of multiple runs. The instantaneous values also demonstrate the superiority of CMT-DA. The other interesting fact we notice is that CMT-DA effectively leverages the path congestion controller and retransmission timeouts to identify the high-congested paths. It helps to alleviate consecutive packet losses and guarantee more successfully received packets.
\section{Conclusion and Discussion}
The future wireless environment is expected to be a converged system that incorporates different access networks with diverse transmission features and capabilities. The increasing powerfulness and popularity of multihomed mobile terminals facilitate the bandwidth aggregation for enhanced transmission reliability and data throughput. Optimizing the SCTP is a critical step towards integrating heterogeneous wireless networks for efficient video delivery.

This paper proposes a novel distortion-aware concurrent multipath transfer (CMT-DA) scheme to support high-quality video streaming over heterogeneous wireless networks. Through modeling and analysis, we have developed solutions for per-path status estimation, congestion window adaption, flow rate allocation, and data retransmission. As future work, we will study the cost minimization problem of utilizing CMT for mobile video delivery in heterogeneous wireless networks.

\section*{Acknowledgement}
\small
This research is partly supported by the Multi-plAtform Game Innovation Centre (MAGIC), funded by the Singapore National Research Foundation under its IDM Futures Funding Initiative and administered by the Interactive $\&$ Digital Media Programme Office, Media Development Authority; National Grand Fundamental Research 973 Program of China under Grant Nos. 2011CB302506, 2013CB329102, 2012CB315802; National High-tech R$\&$D Program of China (863 Program) under Grant No. 2013AA102301; National Natural Science Foundation of China (Grant No. 61132001); Program for New Century Excellent Talents in University (Grant No. NCET-11-0592).

The authors would like to express their sincere gratitude for the anonymous reviewers' helpful comments.


%

\ifCLASSOPTIONcaptionsoff
  \newpage
\fi

\vspace{-22pt}
\begin{IEEEbiography}[{\includegraphics[width=1in,height=1.25in,clip,keepaspectratio]{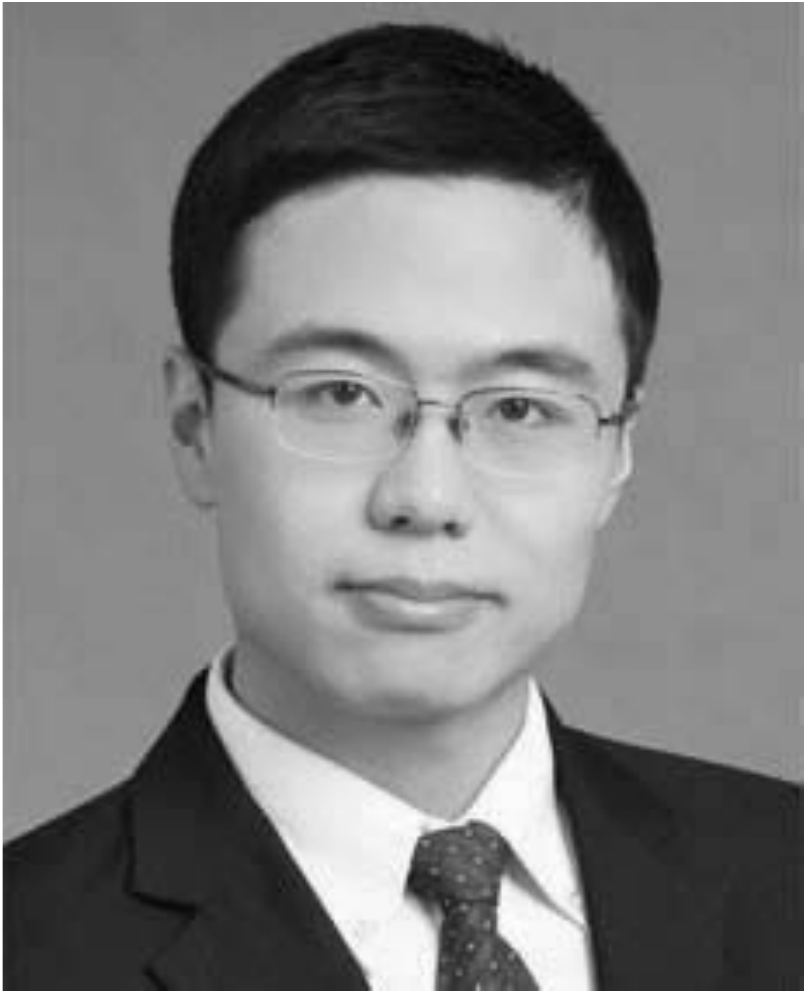}}]{Jiyan Wu} obtained his Ph.D. degree in computer science and technology from the Beijing University of Posts and Telecommunications (supervisor: Prof. Junliang Chen) in June 2014, the Maser degree from China University of Mining and Technology (Beijing) in June 2011. Since March 2014, he has been working as a post-doctoral research fellow (supervisor: Prof. Chau Yuen) in the SUTD-MIT International Design Center, Singapore University of Technology and Design (SUTD). His research interests include video coding, heterogeneous networks, multipath transmission, etc.
\end{IEEEbiography}%
\vspace{-20pt}
\begin{IEEEbiography}[{\includegraphics[width=1in,height=1.25in,clip,keepaspectratio]{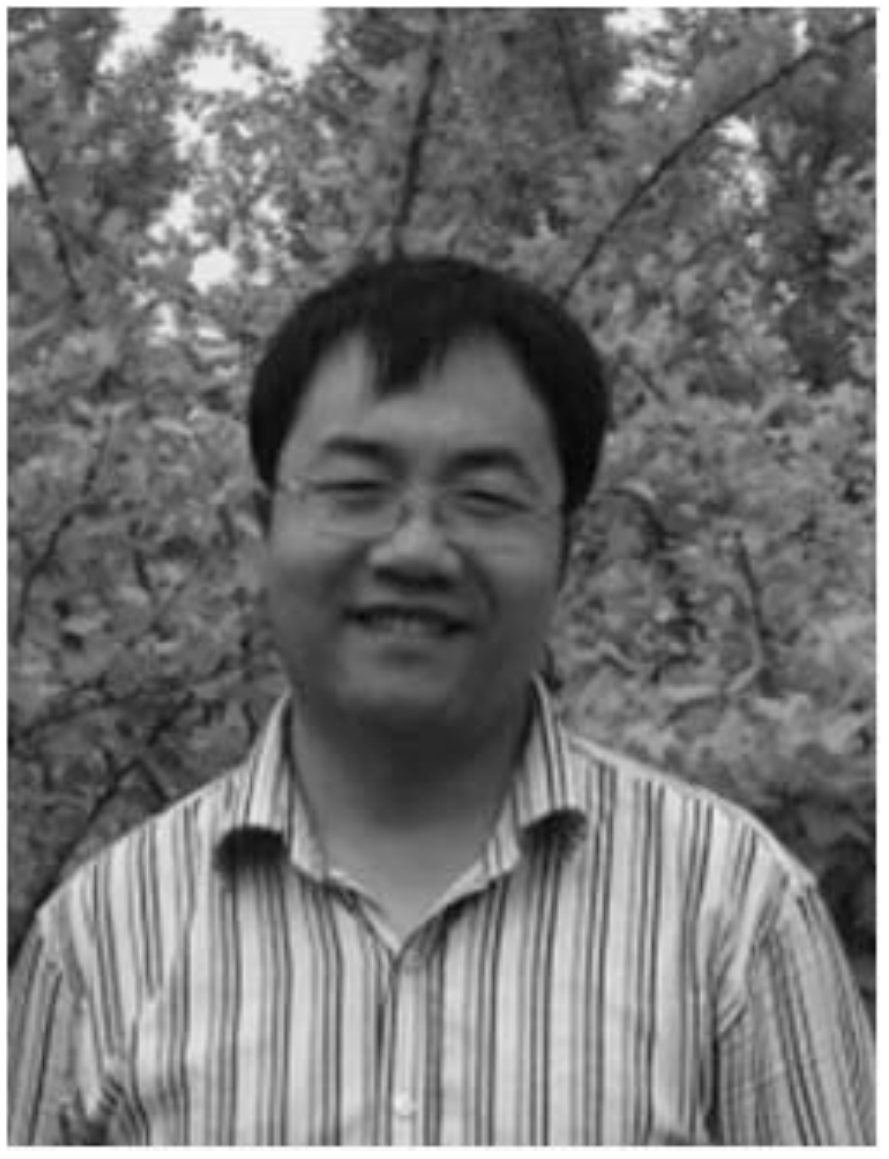}}]{Bo Cheng} received his Ph.D. degree in computer science and engineering in July 2006, from University of Electronic Science and Technology of China. He has been working in the Beijing University of Posts and Telecommunications (BUPT) since 2008. He is now an associate professor of the Research Institute of Networking Technology of BUPT. His current research interests include network services and intelligence, Internet of Things technology, communication software and distribute computing, etc.
\end{IEEEbiography}%
\vspace{-20pt}
\begin{IEEEbiography}[{\includegraphics[width=1in,height=1.25in,clip,keepaspectratio]{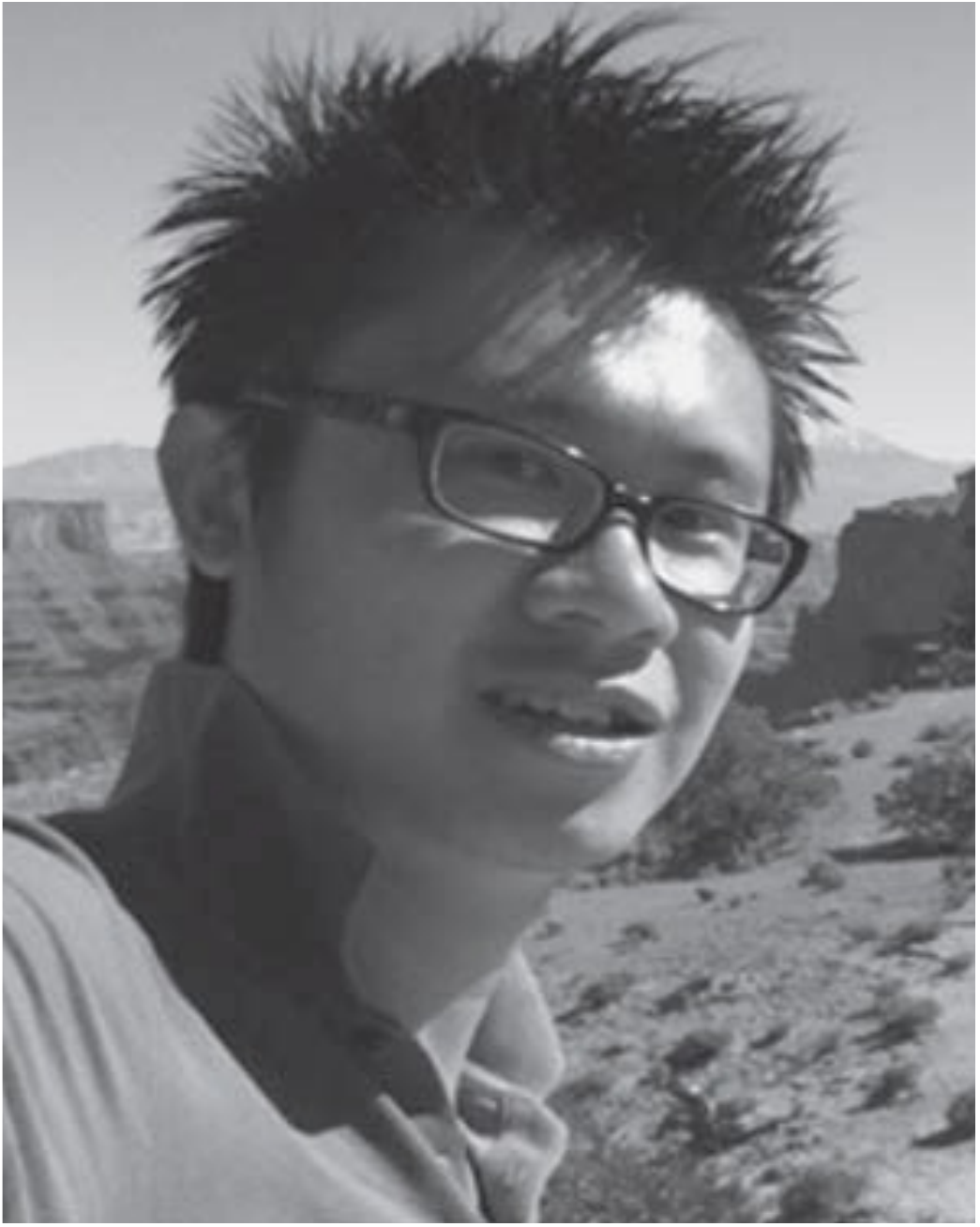}}]{Chau Yuen} received the BEng and PhD degrees from Nanyang Technological University, Singapore, in 2000 and 2004, respectively. He was a postdoc fellow in Lucent Technologies Bell Labs, Murray Hill during 2005. He was also a visiting assistant professor of Hong Kong Polytechnic University in 2008. During the period of 2006-2010, he was at the Institute for Infocomm Research (Singapore) as a senior research engineer. He joined Singapore University of Technology and Design as an assistant professor from June 2010. He also serves as an associate editor for IEEE Transactions on Vehicular Technology. In 2012, he received the IEEE Asia-Pacific Outstanding Young Researcher Award.
\end{IEEEbiography}%
\vspace{-20pt}
\begin{IEEEbiography}[{\includegraphics[width=1in,height=1.25in,clip,keepaspectratio]{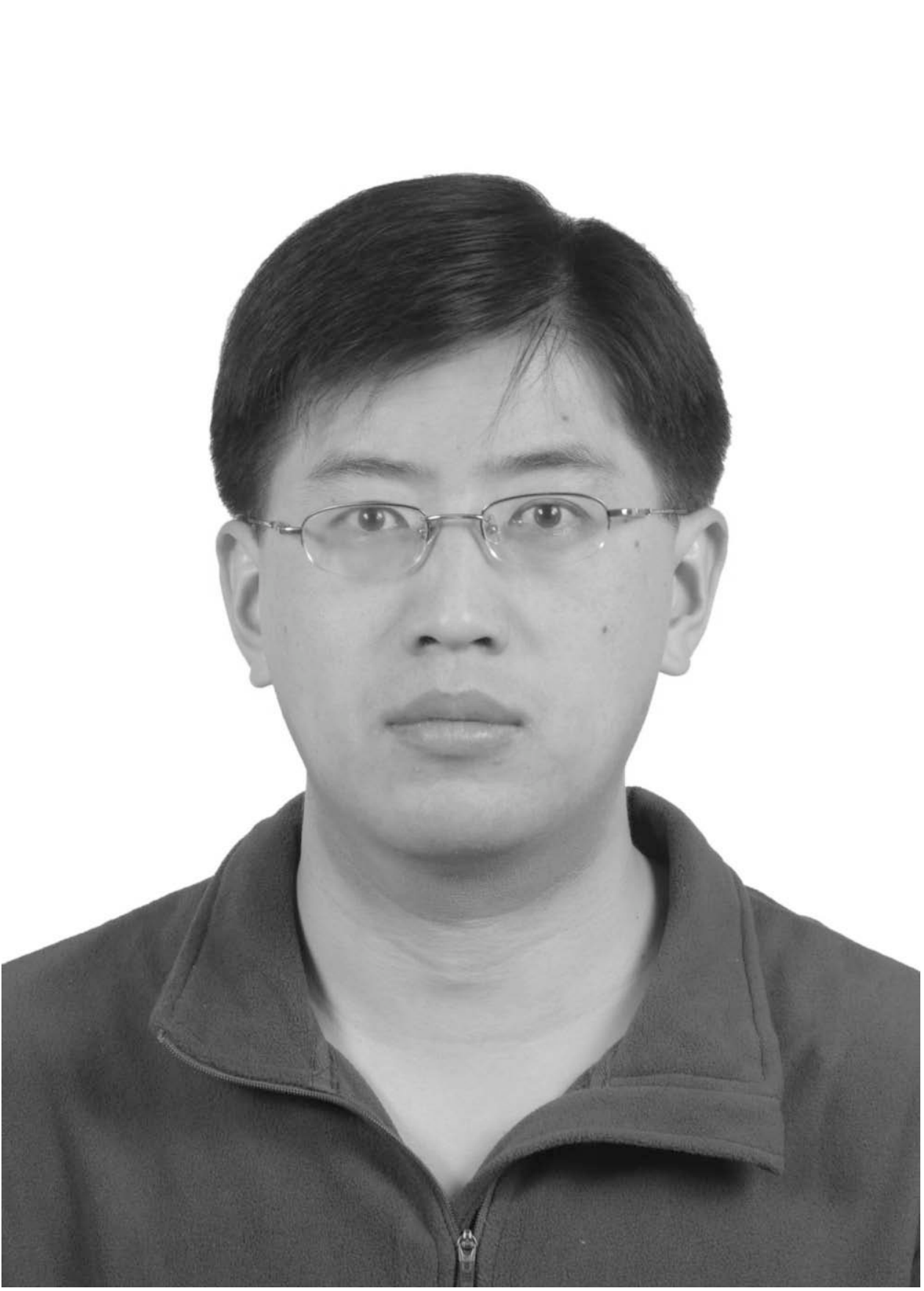}}]{Yanlei Shang} received his Ph.D. degree in computer science and technology from Beijing University of Posts and Telecommunications (BUPT) in 2006. Then he worked in the Nokia Research Center as a postdoctoral research fellow. He is currently an associate professor in the State Key Laboratory of Networking and Switching Technology, BUPT. His research interests include cloud computing, service computing, distributed system and virtualization technology.
\end{IEEEbiography}%
\vspace{-20pt}
\begin{IEEEbiography}[{\includegraphics[width=1in,height=1.25in,clip,keepaspectratio]{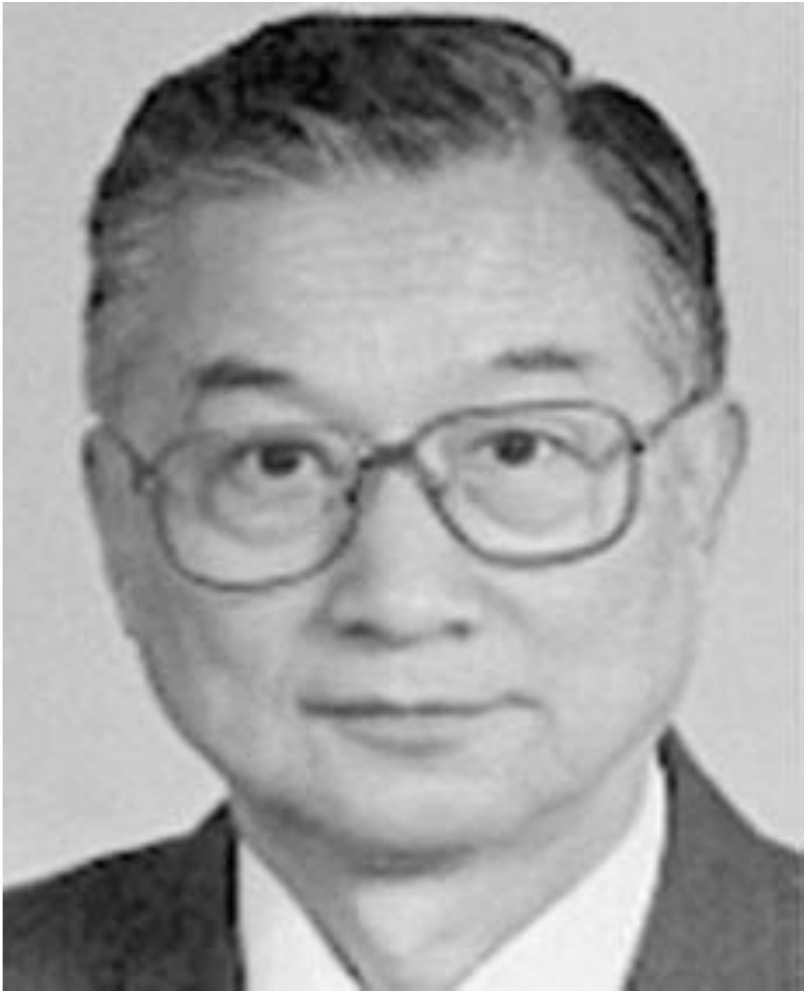}}]{Junliang Chen} is the chairman and a professor of the Research Institute of Networking and Switching Technology at Beijing University of Posts and Telecommunications (BUPT). He has been working in BUPT since 1955. He received the B.S. degree in electrical engineering from Shanghai Jiaotong University, China, in 1955, and his Ph.D. degree in electrical engineering in May, 1961, from Moscow Institute of Radio Engineering, formerly Soviet Russia. His research interests are in the area of communication networks and next generation service creation technology. Prof. Chen was elected as a member of the Chinese Academy of Science in 1991, and a member Chinese Academy of Engineering in 1994.
\end{IEEEbiography}

\end{document}